\documentclass[conference]{IEEEtran}
\IEEEoverridecommandlockouts
\usepackage{cite}
\usepackage{amsmath,amssymb,amsfonts}
\usepackage{algorithmic}
\usepackage{graphicx}
\usepackage{textcomp}
\usepackage{xcolor}
\def\BibTeX{{\rm B\kern-.05em{\sc i\kern-.025em b}\kern-.08em
    T\kern-.1667em\lower.7ex\hbox{E}\kern-.125emX}}

\makeatletter
\newcommand{\linebreakand}{%
  \end{@IEEEauthorhalign}
  \hfill\mbox{}\par
  \mbox{}\hfill\begin{@IEEEauthorhalign}
}
\makeatother

\usepackage[linesnumbered,ruled,vlined]{algorithm2e}
\usepackage{multirow}
\usepackage{pifont}
\usepackage{tikz}
\usepackage{booktabs}
\usepackage{enumitem}
\usepackage{marvosym}
\usepackage{float}
\usepackage{adjustbox}
\usepackage[final]{changes}
\providecommand{\mathadded}[1]{\ifbool{Changes@optiondraft}{\text{\added{\mbox{$\displaystyle #1$}}}}{#1}}
\providecommand{\mathreplaced}[2]{\ifbool{Changes@optiondraft}{\text{\replaced{\mbox{$\displaystyle #1$}}{\mbox{$\displaystyle #2$}}}}{#1}}

\newcommand{\myparatight}[1]{\smallskip\noindent{\bf {#1}.}~}
\newcommand{\graphSize}[1]{\small{#1}}

\newcommand{\reffig}{Figure~}
\newcommand{\reftab}{Table~}

\newtheorem{theorem}{Theorem}
\newtheorem{definition}{Definition}

\newcounter{respAlgo}

\begin{document}

\title{Data Poisoning Attacks to Locally Differentially Private Triangle Counting Protocols
\thanks{This work was supported in part by the National Natural Science Foundation of China (NSFC) under Grants No. 62372226, 62272215, and 62672235.}
}

\author{\IEEEauthorblockN{Tingxuan Han, Wei Tong\IEEEauthorrefmark{1}\thanks{\IEEEauthorrefmark{1}Corresponding author: Wei Tong.}, Zijian Wang, Sheng Zhong}
\IEEEauthorblockA{\textit{State Key Laboratory for Novel Software Technology}, 
\textit{Nanjing University}, 
Nanjing, China \\
hantx.tingxuan@gmail.com, weitong@outlook.com, zj-wang@smail.nju.edu.cn, zhongsheng@nju.edu.cn}
% \and
% \IEEEauthorblockN{Wei Tong}
% \IEEEauthorblockA{\textit{State Key Laboratory for Novel Software Technology} \\
% \textit{Nanjing University}\\
% Nanjing, China \\
% weitong@outlook.com} 
% \linebreakand
% \IEEEauthorblockN{Zijian Wang}
% \IEEEauthorblockA{\textit{State Key Laboratory for Novel Software Technology} \\
% \textit{Nanjing University}\\
% Nanjing, China \\
% zj-wang@smail.nju.edu.cn}
% \and
% \IEEEauthorblockN{Sheng Zhong}
% \IEEEauthorblockA{\textit{State Key Laboratory for Novel Software Technology} \\
% \textit{Nanjing University}\\
% Nanjing, China \\
% zhongsheng@nju.edu.cn}
}

\maketitle

\begin{abstract}
Local differential privacy (LDP) has become a crucial paradigm for privacy-preserving graph analytics, enabling graph analytics tasks such as subgraph counting without requiring a trusted curator. However, while LDP provides strong privacy guarantees, the integrity and robustness of LDP subgraph counting protocols in the presence of malicious participants remain poorly understood. In particular, since data are randomized locally before collection, it is inherently difficult to distinguish malicious users from honest ones based solely on their submitted reports, and small, protocol-compliant local manipulations can be systematically amplified through reconstruction and calibration into large global estimation bias. In this paper, we use LDP triangle counting as a representative task to characterize the vulnerability of LDP subgraph counting to data poisoning attacks. For both non-interactive and interactive LDP triangle counting protocols, we analyze their attack surfaces and develop a unified gain-driven framework that captures how adversarial users can manipulate calibrated subgraph count estimates, which we instantiate on representative protocols of both types. 
%aggregate graph statistics. We further instantiate this framework on representative LDP triangle counting protocols. 
Extensive experiments on four real-world graph datasets demonstrate that these attacks can significantly distort triangle count estimates across different protocols and adversarial settings. Moreover, we show that existing and straightforward defenses are insufficient to mitigate these attacks, underscoring the need for robustness mechanisms tailored to LDP subgraph counting. 

%we evaluate existing defense mechanisms and show that they are insufficient to mitigate these attacks.
\end{abstract}


\section{Introduction}
Local differential privacy (LDP)~\cite{wang2017locally} has become a widely adopted paradigm for privacy-preserving data collection and analysis, particularly in scenarios where users do not trust a centralized curator with their raw data. In graph analytics, LDP enables the collection of sensitive relational information (e.g., user connections or interactions) while providing strong privacy guarantees at the individual level. As a result, LDP has been increasingly applied to fundamental graph analytics tasks such as subgraph counting, which serves as a key building block for higher-level applications~\cite{imola2021locally,ye2020lf,imola2022communication,he2025robust}.

However, while LDP ensures strong privacy guarantees, it may fundamentally weaken the \emph{integrity} of graph analytics. Because each user perturbs their data locally before reporting, the aggregator observes only randomized structural information and cannot verify the correctness of individual reports. 
% This lack of verifiability creates an inherent vulnerability whereby compromised clients can submit \textit{forged} reports with adversarially crafted false-positive or false-negative edges. These poisoned reports satisfy the required format and range constraints, but are designed to bias the aggregator's estimate.
This lack of verifiability creates an inherent vulnerability whereby compromised clients can submit \textit{forged} reports that satisfy the required format and range constraints, yet are intentionally crafted to manipulate the aggregator's estimate. 
% Unlike non-private distributed graph analytics, LDP protocols deliberately introduce false positives and false negatives via local randomization, making adversarial reports harder to distinguish from privacy noise.
More critically, due to structural dependencies in graph data and the reconstruction and calibration procedures commonly used in LDP protocols, such local manipulations can be systematically amplified into large global estimation bias.

Existing robustness studies under LDP have primarily addressed numerical data~\cite{li2023fine,li2024robustness}, categorical data~\cite{cheu2021manipulation,cao2021data,huang2024ldpguard, tong2024data,song2023efficient}, or key-valued data~\cite{wu2022poisoning, horigome2023local}. Graph data pose additional challenges due to their relational structure and correlations, which allow adversaries to induce coordinated and amplified effects. While recent work has explored attacks on node-level graph metrics~\cite{he2025data}, which target per-user estimates, the robustness of subgraph counting, particularly triangle counting, remains largely unexplored.

% \begin{figure}[ht]
% % \tabcolsep=0.1cm
% % \centering
% \begin{tabular}{cc}
% \includegraphics[width=0.2\textwidth]{figures/example_a.pdf}
% &\includegraphics[width=0.2\textwidth]{figures/example_b.pdf}\\
% \graphSize{(a) Global graph}
% &\graphSize{(b) Subgraph counts}\\
% \end{tabular}
% \caption{Example of subgraph counts}
% \label{fig:example}
% \end{figure}

In this work, we investigate data poisoning attacks against LDP triangle counting, one of the most fundamental subgraph counting tasks. We propose an attack framework applicable to both non-interactive and interactive LDP protocols, and instantiate concrete attack strategies for different protocols. Our attacks demonstrate how an adversary can craft poisoned reports from compromised users to drive the calibrated triangle count toward attacker-chosen targets, thereby amplifying estimation bias and misleading downstream inferences about global graph structure. 
Such manipulation may have practical consequences, for example, by degrading ad targeting, steering malicious traffic, or triggering resource misallocation.

Designing effective poisoning attacks in this setting is challenging due to the interplay between local perturbation and global structural effects. First, graph data exhibit inherent structural complexity that differs significantly from numerical, categorical, or key–value data, as nodes and edges are tightly coupled and operations on different nodes are not independent but mutually influence each other through the graph structure. Second, compounding this structural complexity, LDP perturbation mechanisms play a qualitatively different role in graph analysis. Unlike many LDP protocols that directly rely on mechanisms such as GRR~\cite{kairouz2014extremal}, OLH~\cite{wang2017locally}, or OUE~\cite{wang2017locally} for perturbation and aggregation, LDP graph protocols typically use Randomized Response (RR)~\cite{warner1965randomized} to collect local structural information, followed by structure-dependent calibration and estimation procedures. As a result, effective poisoning must jointly exploit both the perturbation mechanism and the graph's structural properties. Third, beyond these per-user challenges, poisoning subgraph counting targets a \emph{global} outcome involving all users, which fundamentally differs from attacks that manipulate estimates for a small set of target users in node-level metric estimation~\cite{he2025data}. The expanded attack scope, together with the fundamental differences between node-level graph metric estimation and subgraph counting, necessitate attack strategies specifically tailored to LDP subgraph counting tasks. 

To address these challenges, we first define a threat model and identify the attack surface for LDP triangle counting protocols. Building on this formulation, we propose two randomized attack baselines for triangle counting, RPA$_{\triangle}$ and RTA$_{\triangle}$, which perturb reported data and true local data, respectively. We then develop a unified gain-driven greedy poisoning framework, DPA$_{\triangle}$, that applies to both non-interactive and interactive protocols. In the non-interactive setting, we instantiate our attacks against LocalRR$_{\triangle}$~\cite{imola2021locally} and TCountRR~\cite{eden2025triangle}, which together represent the two main calibration paradigms: empirical estimation from global pattern counts and per-edge unbiased calibration. In the interactive setting, we target ARROneNS$_{\triangle}$~\cite{imola2022communication}, TDP-VC~\cite{he2025robust}, and EdgeOrient$_{\triangle}$~\cite{mundra2025practical}, which span different strategies for leveraging local structural information in the second round of interaction.

We first study a full-knowledge adversary and then extend our attacks to a more realistic partial-knowledge setting that relies only on perturbed data and the local views of compromised users. Through extensive empirical evaluation on four real-world graph datasets, we show that the proposed attack strategies can substantially degrade the accuracy of LDP triangle counting across various protocols and adversarial settings. 
Moreover, since defenses developed for standard LDP mechanisms do not readily extend to graph analytics, we examine two natural aggregator-side mitigation strategies and show that they are insufficient, highlighting the need for robustness mechanisms tailored to LDP subgraph counting.
% we propose two defense schemes and evaluate their mitigation effects, highlighting the need for robustness mechanisms tailored to LDP-based graph analytics. 

Our contributions are summarised as follows: 
\begin{itemize}
    \item To the best of our knowledge, we provide the first systematic study of data poisoning attacks against LDP triangle counting protocols, revealing that the interplay between local randomization, graph structure, and protocol-level calibration creates an attack surface qualitatively different from that of existing LDP poisoning settings.

    \item We formulate data poisoning in LDP triangle counting as a gain-driven optimization problem over graph structure and develop a unified gain-driven greedy poisoning framework (DPA$_{\triangle}$). The framework shares a single estimator-driven selection principle across protocols, which we instantiate on representative non-interactive and interactive protocols.

    \item We conduct extensive experiments on four real-world datasets with two baseline attacks, evaluating our methods under both full-knowledge and partial-knowledge adversaries.
    
    \item We discuss and evaluate potential mitigation solutions, showing that existing and straightforward defenses are either inapplicable or insufficiently effective.
\end{itemize}

\section{Background}
\subsection{Local Differential Privacy for Graphs}
Differential privacy~\cite{dwork2006differential} is a widely adopted standard for formalizing privacy guarantees in data analysis. Local differential privacy ~\cite{wang2017locally} is the local model of DP, where each user randomizes her own data before sending a report to an untrusted aggregator. 
% We recall the formal definition of local differential privacy.

% \begin{definition}[Local Differential Privacy]
%     For $\epsilon\in \mathbb{R}_{\ge 0}$, a randomized mechanism $\mathcal{M}$ satisfies $\epsilon$-LDP if for any two neighboring inputs $x$ and $x'$ in its domain and any subset of outputs $O\in Range(\mathcal{M})$, we have
% \begin{equation}
%     \Pr[\mathcal{M}(x) \in O] \le e^{\epsilon} \Pr[\mathcal{M}(x')\in O].
% \end{equation}
% \end{definition}

LDP is now widely used for collecting and analyzing binary~\cite{erlingsson2014rappor}, key-value~\cite {ye2019privkv}, location~\cite{errounda2021collective}, and graph data~\cite{ye2020lf}. We consider a simple undirected graph $G=(V,E)$ with $V = \{v_1, v_2, \dots,v_n\}$ and edges $E \subseteq \{\{u, v\}: u, v\in V,\, u \neq v\}$.
In graph analytics under LDP, privacy is specified through a neighboring relation that captures what constitutes a single individual's sensitive change. Two common notions are node local differential privacy (node LDP)~\cite{kasiviswanathan2013analyzing} and edge local differential privacy (edge LDP)~\cite{qin2017generating}. Under node LDP, two neighboring graphs differ by the presence of one node together with all its incident edges. Whereas two neighboring graphs differ by a single edge under edge LDP. Since the LDP triangle counting protocols targeted in this paper are designed under the edge level, we adopt edge LDP as the privacy standard.

\begin{definition}[Edge Local Differential Privacy]
    For $\epsilon \in \mathbb{R}_{\ge 0}$, a randomized mechanism $\mathcal{M}$ satisfies $\epsilon$-edge LDP if for any two neighboring inputs $x$ and $x'$ that differ in exactly one edge and any subset of outputs $O\in Range(\mathcal{M})$, we have
\begin{equation}
    \Pr[\mathcal{M}(x)\in O] \le e^{\epsilon} \Pr[\mathcal{M}(x')\in O].
\end{equation}
\end{definition}

% \begin{table}[htbp]
%   \caption{Notations}
%   \label{tab:notation}
%   \small
%   \begin{tabular}{c|l}
%     \toprule
%     Notation &Description \\
%     \midrule
%     $G, G', G_p'$ 
%     &Original graph, perturbed graph, poisoned graph\\
%     $v_i$ 
%     &The $i$-th node in $G$\\
%     $G_i$ 
%     &Local view of $v_i$\\
%     $V, V_i$ 
%     &Set of nodes contained by $G, G_i$\\
%     $E,E_i,E',E_p$ 
%     &Set of edges contained by $G,G_i,G',G'_p$\\ 
%     $n,n_c,n_h$
%     &Numbers of total, compromised, honest nodes\\
%     $V_c$
%     &Set of compromised nodes in $G$\\
%     $c$
%     &Attack capability of each compromised node\\
%     $y_i,\hat{y}_i$
%     &Honest and poisoned report under LDP protocols \\
%     $Y_i,\hat{Y}_i$
%     &Set of honest and poisoned reports\\
%     $\epsilon,\epsilon_1,\epsilon_2$
%     & Privacy budgets\\
%   \bottomrule
% \end{tabular}
% \end{table}

\subsection{Triangle Counting under LDP}
A triangle is a 3-cycle: nodes $v_0, v_1, v_2$ form a triangle if $\{v_0,v_1\}$, $\{v_1,v_2\}$, $\{v_2,v_0\}\in E$. Formally, the function for triangle counting can be formalized as $f_{\triangle}(G)$: $\mathcal{G}_n \to \mathbb{Z}_{\ge 0}$. For LDP triangle counting protocols, we categorize existing approaches into two types: non-interactive and interactive protocols. In non-interactive protocols, the aggregator collects users’ neighboring lists after they have been perturbed via Randomized Response mechanism, reconstructs a perturbed graph from these perturbed reports, and then estimates the number of triangles with an additional calibrating step.
By contrast, interactive protocols introduce multiple rounds of communication between the aggregator and the users. In this setting, the aggregator can send the perturbed graph information back to each user. This interaction helps compensate for the fact that an individual user cannot locally determine whether the third edge of a triangle exists among its neighbors, and therefore cannot accurately compute the number of triangles incident to itself using only local information. As a result, interactive protocols enable more complete utilization of each user’s true local information, thereby improving the accuracy of triangle counting.

\subsubsection{Non-interactive Protocols}
Non-interactive LDP triangle counting protocols require each user to perturb its local structural information once and submit a single report to the aggregator. The aggregator then reconstructs a noisy graph or calibrated edge statistics and estimates the triangle count by applying a post-processing calibration step. We briefly introduce two representative protocols, LocalRR$_{\triangle}$~\cite{imola2021locally} and TCountRR~\cite{eden2025triangle}.

\myparatight{LocalRR$_{\triangle}$}
Each user encodes its local neighbor list as a binary vector and applies Randomized Response to each required entry. The aggregator combines all randomized reports to construct a perturbed graph $G'$. It then counts the four possible three-node patterns in $G'$, denoted by $m_3,m_2,m_1,m_0$ for triangles, 2-stars, 1-edge, and no-edge patterns, respectively. The final triangle estimate is obtained by applying an empirical calibration formula to these pattern counts. 

\myparatight{TCountRR}
TCountRR also uses Randomized Response to perturb local edge indicators, but it calibrates each perturbed edge separately before triangle aggregation. For each reported edge variable $X_{i,j}$, the aggregator constructs an unbiased edge estimator
\begin{equation}
\begin{aligned}
    Y_{i,j}
    =
    \frac{X_{i,j}(e^{\epsilon}+1)-1}{e^{\epsilon}-1}.
\end{aligned}
\end{equation}
The triangle count is then estimated by aggregating products of calibrated edge variables over all node triples:
\begin{equation}
\begin{aligned}
    \hat{T}
    =
    \sum_{i<j<k}Y_{i,j}Y_{i,k}Y_{j,k}.
\end{aligned}
\end{equation}
% Therefore, TCountRR represents a \emph{per-edge unbiased calibration} strategy.

\subsubsection{Interactive Protocols}
Interactive LDP triangle counting protocols reduce estimation noise by allowing the aggregator to send perturbed structural information back to users. Each user can then combine its true local structure, such as local 2-stars, with noisy information about closing edges and submit a locally computed statistic. Representative protocols include ARROneNS$_{\triangle}$~\cite{imola2022communication}, TDP-VC~\cite{he2025robust}, and EdgeOrient$_{\triangle}$~\cite{mundra2025practical}.

\myparatight{ARROneNS$_{\triangle}$}
In the first round, users perturb their neighbor lists and send the reports to the aggregator. The aggregator distributes selected noisy graph information back to users. In the second round, each user combines its true local 2-stars with the received noisy edge information, adds calibrated noise, and submits a perturbed triangle counting report. The aggregator sums these reports to obtain the final triangle estimate.

\myparatight{TDP-VC}
TDP-VC further improves the interactive framework by using user-specific sensitivity control. Each user perturbs its degree and neighbor list, and the aggregator constructs a perturbed graph. In the second round, each user samples a local neighborhood according to its noisy degree and estimates triangle contributions using calibrated noisy edges among the sampled neighbors. The final estimate is obtained by aggregating users' perturbed reports.

\myparatight{EdgeOrient$_{\triangle}$}
EdgeOrient$_{\triangle}$ reduces the number of noisy edges involved in triangle estimation by first computing a low-out-degree ordering under LDP. Given this ordering, each user only considers eligible neighbor pairs that satisfy the orientation constraint. The user then combines its local structure with perturbed structural information and submits a noisy local triangle estimate. The aggregator sums all perturbed estimates to obtain the final count.

Overall, non-interactive protocols estimate triangles directly from perturbed reports, whereas interactive protocols estimate local triangle contributions by combining true local wedges with noisy closing-edge information. This distinction is important for our attack analysis, because poisoned reports affect the former through reconstructed graph statistics and the latter through local wedge-closing contributions.

\section{Attack Model and Formulation}
\label{sec:attack_formulation}
\replaced{We formulate attacks that modify compromised users' reports to construct a poisoned graph $G'_p$ and bias triangle estimation through reconstruction, interaction, and calibration. We specify the threat model, attack surface, goal, and baseline strategies.}{In this section, we formalize data poisoning attacks against locally differentially private triangle counting protocols. By manipulating reports from a subset of compromised users, the attacker induces the aggregator to construct a perturbed, poisoned graph $G'_p$. This poisoned graph $G'_p$ propagates through the protocol's interaction, aggregation, and calibration steps, ultimately yielding a biased estimate of the triangle count. Concretely, we present our formulation from four aspects: the threat model, the attack surface, the attack goal, and baseline attack strategies.}

\subsection{Threat Model}
\label{sec:threat_model}
We describe our threat model in terms of the attacker's background knowledge and capabilities.

\myparatight{Attacker’s background knowledge} We assume the attacker is familiar with the protocol's details, including local randomizers and privacy parameters. Depending on how much the attacker knows about the original graph $G=(V, E)$ and the perturbed graph $G'=(V, E')$, we consider two levels of background knowledge,  \emph{full knowledge} and \emph{partial knowledge}, as commonly adopted in prior work on poisoning attacks against LDP protocols~\cite{cao2021data,li2023fine,tong2024data}.

% \begin{figure}[ht]
% % \tabcolsep=0.1cm
% % \centering
% \begin{tabular}{cc}
% \includegraphics[width=0.2\textwidth]{figures/example_c.pdf}
% &\includegraphics[width=0.2\textwidth]{figures/example_d.pdf}\\
% \graphSize{(a) $v_1$'s $1$-hop view}
% &\graphSize{(b) $v_1$'s $2$-hop view}\\
% \end{tabular}
% \caption{Example of $v_1$'s $s$-hop view}
% \label{fig:extended_view}
% \end{figure}

\begin{itemize}
\item \emph{Full knowledge.} The attacker possesses all information of the graphs $G$ and $G'$, including all nodes and edges. \replaced{This strong-knowledge setting serves as a reference for evaluating how restricting graph information affects attack performance. The attacker remains subject to the same poisoning budget and reporting permissions under both knowledge settings.}{This is a strong but common assumption used to characterise the upper bound of attack effectiveness.}

\item \emph{Partial knowledge.} The attacker knows \deleted{only} the local views of the compromised users. In addition, the attacker may observe the raw graph data of some honest users (e.g., by observing neighbors' connections) or the noisy reports received by the aggregator (e.g., by eavesdropping on the communication channel)\cite{cheng2025atom,wu2022model}. \replaced{These observations depend on the available side information and communication interface. Their availability is an explicit assumption of the attack setting and is not guaranteed by LDP. Section~\ref{sec:partial_knowledge} specifies the true and perturbed information available for each protocol and knowledge level.}{This setting captures the background knowledge that an attacker may obtain in the real world. The attacker’s partial knowledge is constrained by the original graph topology. That means it can access only the true or perturbed information within its true $s$-hop neighborhood.} Formally, the $s$-hop neighborhood of a node $u$ is the set of all nodes that can be reached from $u$ by traversing at most $s$ edges.
\end{itemize}

\myparatight{Attacker's capabilities}
\label{sec:capability}
\replaced{Following prior work~\cite{cao2021data,cheu2021manipulation,he2025data}, we assume the attacker can compromise a subset of users and replace their reports with forged reports that satisfy the protocol's format and value-range requirements. Let $V_c\subseteq V$ denote the compromised users, with $n_c=|V_c|$ and $n_h=n-n_c$. Each compromised user may make at most $c$ local adjacency edits~\cite{wang2019attacking}. For DPA$_\triangle$ and RPA$_\triangle$, edits are counted relative to the perturbed report before poisoning. For RTA$_\triangle$, they are counted relative to the true local input before randomization. Each insertion or deletion counts as one edit, and only entries assigned to the compromised user may be changed. The original graph $G$ is unchanged.}{Following prior work~\mbox{\cite{cao2021data,cheu2021manipulation,he2025data}}, we assume the attacker can compromise a subset of users participating in the decentralized triangle counting protocol and replace their reports sent to the aggregator with forged reports generated following the protocol. Let $V_c\subseteq V$ be the set of compromised users, with $n_c=|V_c|$, and let $n_h=n-n_c$ denote the number of honest users. To model different poisoning strengths and stealthiness levels, we further assume that each compromised user has an attack capability $c$, which bounds the maximum number of \textit{local edge modifications} allowed in its reported view\mbox{\cite{wang2019attacking}}. Specifically, for each compromised user $v_i\in V_c$, the attacker can construct a \textit{poisoned local view} (e.g., a neighbor list or an adjacency vector) that differs from user $i$'s true local view by at most $c$ edge modifications. The attacker only manipulates submitted data under the LDP protocols and does not modify the original graph $G$.}

\added{The parameters $\beta=n_c/n$ and $c$ describe attacker coverage and per-user manipulation strength. We vary them separately to distinguish attacks involving more users from attacks making more changes to each report. }

\myparatight{\added{Auditing model}}
\added{The aggregator observes protocol parameters, participant identities, reporting assignments, and submitted reports. It can check format, dimensions, ownership, and value ranges. Without trusted pre-poisoning inputs or reports, it cannot directly verify the edit budget $c$. Passing these checks does not imply statistical undetectability. Section~\ref{sec:defense} evaluates degree-based pruning and frequent-pattern detection.}

\subsection{Attack Surface}
\label{sec:attac_surface}
The attack surface lies in the randomized adjacency information reported by users, typically in the form of a perturbed neighbor list or adjacency vector. Based on these reports, the aggregator reconstructs a perturbed graph $G'$ and then estimates triangle count, followed by aggregation and calibration. In this setting, poisoning is realized by modifying a small number of adjacency entries in the reported structure while still preserving protocol-compliant reports. Such seemingly minor structural edits can have a large effect, as they may alter the reconstructed degrees of multiple nodes and simultaneously close many 2-stars into triangles in the reconstructed graph, thereby amplifying the final estimation error. This makes structural reports a particularly effective attack surface, as local manipulations can propagate through both the reconstruction and calibration stages of the protocol. Accordingly, our attacks are designed as feasible local edits that exploit this structural leverage while remaining consistent with the protocol interface.

\subsection{Attack Goal}
\label{sec:attack_goal}
The attacker aims to degrade the accuracy of triangle counting under LDP, thereby misleading the aggregator and harming downstream graph analytics (e.g., community detection, link prediction). Because the attacker aims to manipulate the \textit{global} triangle estimate, every user is considered a potential target. For each user $v_i \in V$, let $y_i$ be the report that $v_i$ would submit under the LDP protocol if honest, and let $\hat{y}_i$ be the poisoned report submitted when $v_i \in V_c$ is compromised. 
Let $Y = \{y_1, y_2, \dots, y_n\}$ be the collection of honest reports, and let $\hat{Y}$ be obtained by replacing $y_i$ with $\hat{y}_i$ for all $v_i \in V_c$. 
Let $f(\cdot)$ be the protocol’s estimator for triangle count. We define the attacker’s gain as
\begin{equation}
    \begin{aligned}
        \mathrm{Gain} = \left|f(Y) - f(\hat{Y})\right|,
    \end{aligned}
\end{equation}
which measures the impact of the data poisoning attack. The attacker aims to choose poisoned reports $\hat{Y}$, constrained by the LDP protocol to maximize this gain: 
\begin{equation}
    \begin{aligned}
        \max_{\hat{Y}} \; \mathrm{Gain} = \max_{\hat{Y}} \; |f(Y) - f(\hat{Y})|.
    \end{aligned}
\end{equation}
Since the deviation can be increased by either inflating or deflating the estimate, without loss of generality, we focus on attacks that induce a positive bias.

\subsection{Baseline Attacks}
\label{sec:baselines}
To the best of our knowledge, there have been no prior studies on poisoning LDP triangle counting. Inspired by random attack strategies in related domains~\cite{cao2021data, he2025data}, we introduce two simple baselines for comparison: \emph{Random Perturbed-Report Attack (RPA${_\triangle}$)} and \emph{Random True-Value Attack (RTA${_\triangle}$)}. These two attacks serve as baselines for evaluating the extent of improvement our approach can achieve over uninformed random manipulation. 

\myparatight{Random Perturbed-Report Attack (RPA${_\triangle}$)} This attack operates \emph{after} the LDP mechanism is applied. Each compromised user first generates a protocol-compliant perturbed report as usual, after which the attacker replaces it with a random but still \emph{valid} report. Specifically, for each compromised user, randomly selecting up to $c$ non-edges in its perturbed neighbor list and flipping them to edges, and then submitting the modified perturbed report to the aggregator.

\myparatight{Random True-Value Attack (RTA${_\triangle}$)} This attack operates \emph{before} the LDP mechanism is applied. For each compromised user, the attacker first modifies the user's local input within the capability budget to obtain a poisoned local view, and the user then applies the LDP protocol to this poisoned input before submission. Specifically, the attacker randomly rewires up to $c$ incident edges in the user’s \emph{true} neighbor list and then runs the LDP mechanism on the resulting poisoned local view.

\section{Poisoning LDP Triangle Counting Protocols}
We consider two common protocol patterns for LDP triangle counting. 
In \emph{non-interactive} protocols, each user perturbs its adjacency information once and sends a single perturbed report to the aggregator, which estimates triangles based on these perturbed reports (e.g., LocalRR$_\triangle$~\cite{imola2021locally}, TCountRR~\cite{eden2025triangle}). 
While \emph{interactive} protocols typically contain a two-round interaction as the key operation. In the first round, each user perturbs its local information and sends a perturbed report to the aggregator. In the second round, the aggregator broadcasts the first-round perturbed information, and each user combines this broadcast message with its local ground truth to compute a triangle-related statistic, perturbs it again as specified, and sends it back for final aggregation and calibration (e.g., ARROneNS$_{\triangle}$~\cite{imola2022communication}, TDP-VC~\cite{he2025robust}, EdgeOrient$_{\triangle}$~\cite{mundra2025practical}).

\subsection{Attack Framework}
For triangle counting, the attacker aims to increase the final estimate $f_{\triangle}$ by manipulating the perturbed reports submitted by compromised users. These forged reports effectively modify the privatized view (e.g., the reconstructed perturbed graph) on which the protocol performs triangle estimation and calibration, thereby shifting the final output in the attacker-chosen direction.

% \begin{figure}[ht]
% \centering
% \includegraphics[width=0.3\textwidth]{figures/triangle.pdf}
% \caption{\MajorColor{Types of new triangles}}
% \label{fig:triangle}
% \end{figure}

To reason about the combinatorial impact of edge insertions, we classify newly created triangles by the number of poisoned edges they contain: $\mathrm{triangle}_j$ denotes the number of new triangles that contain exactly $j$ poisoned edges, for $j \in \{1,2,3\}$.
Optimizing the total number of new triangles subject to a budget on edge insertions is computationally intractable in general~\cite{dehghani2015using}.
Motivated by this hardness, we adopt a greedy, gain-driven poisoning framework (Algorithm~\ref{alg:DPATC}) that serves as a unified template across triangle counting protocols.

\begin{algorithm} [t]
    \SetKwFunction{Gain}{Gain}  
    \SetKwFunction{Update}{Update} 
    \SetKwFunction{Degree}{Deg}
    \SetKwData{GainMtx}{GainMtx} 
    \SetKwData{CapList}{CapList}
    \SetKwInOut{Input}{Input}
    \SetKwInOut{Output}{Output} 
    
    \Input{Graph $G=(V,E)$, perturbed graph $G'=(V,E')$, \\ compromised users $V_c$, poisoning capability $c$}
    \Output{Poisoned graph $G'_p=(V,E'\cup E_p)$}
    \BlankLine 
    Initialize $E_p = \emptyset$\;
    Initialize \GainMtx as an $(n \times n)$-dimensional matrix\;
    Initialize \CapList as an $n$-dimensional list\;
    \For{$u \in V_c$}{
        \CapList$[u] \gets c$\;
        \For{$\{u, v\} \notin E'$}{
            \GainMtx$[u,v] \gets $ \Gain{$u,v$}\;
        }
    }
    \While{$\sum_{u\in V_c}\CapList[u] > 0$}{
        $u, v \gets \arg\max_{u\in V_c,\, v\in V}\;\GainMtx[u,v]$\;
        \If{\GainMtx$[u,v] \le 0$}{
            $v \gets \arg\max_{x\in V}\;\Degree(x, E'\cup E_p)$\;
            $u \gets \arg\max_{x\in V_c \setminus N_{E'\cup E_p}(v)}\;\Degree(x, E'\cup E_p)$\;
        }
        $E_p \gets E_p \cup \{uv\}$\;
        \CapList$[u] \gets \CapList[u]-1$\;
    $V_c, \GainMtx \gets \Update(G'_p, V_c, \GainMtx, \CapList)$\;
    }
    \Return{$G'_p=(V,E'\cup E_p)$}\;
    \caption{Positive-Bias Data Poisoning Attack on Triangle Counting}\label{alg:DPATC} 
\end{algorithm}

\label{app:algorithms}
\begin{algorithm} [t]
    \SetKwFunction{Gain}{Gain}  \SetKwFunction{Update}{Update}
    \SetKwData{GainMtx}{GainMtx} \SetKwData{CapList}{CapList}
    \SetKwInOut{Input}{Input}\SetKwInOut{Output}{Output} 
    
    \Input{Poisoned graph $G_p'=(V,E'\cup E_p)$, \\ set of compromised users $V_c$, gain matrix \GainMtx, \\ list of poisoning capability \CapList}
    \Output{Set of remaining compromised users $V_c$, \\
    gain matrix \GainMtx}
    \BlankLine 
    % \tcp{[$i$]/[$a$] indicates the process is executed by party $i$/the aggregator.}
    \For{$u \in V_c$}{
        \If{\CapList$[u]\le 0$}{
             \GainMtx$[u,:] = 0$\;
             $V_c=V_c-\{u\}$\;
             \textbf{break}\;
        }
        \For{$uv\notin E'\cup E_p$}{
            \GainMtx$[u,v] =$ \Gain{$u,v$}\;
        }
    }
    \Return{$V_c$, \GainMtx\;}
    \caption{\texttt{Update}}\label{alg:Update} 
\end{algorithm}

At a high level, Algorithm~\ref{alg:DPATC} iteratively selects a feasible poisoned edge to insert under per-user capability constraints.
Each candidate edge is scored by a \emph{protocol-dependent gain} that measures its marginal increase in the positive-bias attack objective, i.e., the calibrated triangle estimate.
When no positive-gain edge exists under the current poisoned view (e.g., no currently closable 2-star exists), the insertion attack performs a seeding step to create future opportunities. The rationale is that an inserted edge may have zero immediate gain if it does not close any existing wedge, but it can create new wedges that later poisoned edges may close. Therefore, the attacker connects a maximum-degree vertex to a high-degree compromised vertex that is currently nonadjacent.
The framework consists of three steps: \emph{Initialization}, \emph{Gain-driven Edge Selection}, and \emph{Gain Update}.

\myparatight{Initialization} The attacker initializes the set of poisoned edges $E_p$, the gain matrix \textsf{\small GainMtx}, and the capability list \textsf{\small CapList}, which records the remaining poisoning budget of each compromised user. 
Here, \textsf{\small GainMtx}$[u,v]$ represents the marginal increase in the positive-bias attack objective if the compromised user $u$ inserts the candidate edge $\{u,v\}$.
Edges that already exist in the current view yield no marginal gain, and edges between two honest users are infeasible to modify. Thus, we set the corresponding entries in \textsf{\small GainMtx} to zero.

\myparatight{Gain-driven Edge Selection} Edges are iteratively selected until all poisoning capabilities are exhausted. At each step:
\begin{enumerate}
    \item If the maximum entry in \textsf{\small GainMtx} is positive, select the edge $\{u, v\}$ with the largest gain.
    \item Otherwise, no positive-gain insertion exists under the current poisoned view (e.g., no closable 2-star). We then perform the seeding step: choose a node $v$ with the maximum degree in $G_p$, and choose a compromised user $u \in V_c$ with remaining poisoning capacity that is not adjacent to $v$ in $G_p$ and has the largest degree among such candidates. Add $\{u,v\}$ as the next poisoned edge.
\end{enumerate}

\myparatight{Gain Update} After an edge is added, we decrement \textsf{\small CapList}$[u]$ by one and update $V_c$ and \textsf{\small GainMtx} via two operations. 
First, we set to zero the \textsf{\small GainMtx} entries associated with any compromised user whose poisoning capacity is exhausted and remove that user from $V_c$. 
Second, we update the relevant entries of \textsf{\small GainMtx} for the remaining compromised users in $V_c$ based on the current poisoned graph $G_p'$. Details are provided in Algorithm~\ref{alg:Update}. 

Since the output after calibration is the final estimate produced by the LDP protocol, designing the gain function based on the calibration step and exploiting the users' reporting stage to poison the final estimate are key factors that determine the effectiveness of the attack. Before instantiating the gain functions for concrete protocols, we first clarify how the framework supports both positive-bias and negative-bias objectives and analyze its privacy and complexity properties.

\subsection{Positive-Bias and Negative-Bias Attacks}
\label{sec:pos_neg_attack}

The framework above can be instantiated in two directions of attack. In a \emph{positive-bias attack}, the attacker aims to increase the final estimate of the triangle counting. This is achieved by inserting poisoned reported edges, so that additional wedges are closed into triangles after reconstruction or during user-level local counting. In a \emph{negative-bias attack}, the attacker aims to decrease the final estimate. This is achieved by deleting existing reported edges controlled by compromised users, thereby removing triangles or triangle-related local contributions from the estimator.

Formally, let $f(\cdot)$ denote the triangle counting estimator for the target protocol. For a positive-bias insertion candidate $e\notin E'_p$, the insertion gain is defined as
\begin{equation}
    \mathrm{Gain}^{+}(e)
    =
    f(G'_p\cup\{e\})-f(G'_p).
\end{equation}
The attacker greedily inserts feasible candidate edges with positive $\mathrm{Gain}^{+}$. For a negative-bias deletion candidate $e\in E'_p$, the deletion gain is defined as the decrease in the calibrated estimator after removing $e$:
\begin{equation}
    \mathrm{Gain}^{-}(e)
    =
    f(G'_p)-f(G'_p\setminus\{e\}).
\end{equation}
The attacker greedily deletes feasible candidate edges with positive $\mathrm{Gain}^{-}$ until the poisoning budget is exhausted or no positive-gain deletion remains.

The two directions share the same estimator-driven greedy principle, but they differ in their feasible candidate spaces. Insertion attacks choose from non-existing reported edges that compromised users can add. This candidate space is usually large, and an inserted edge can create new wedge-closing opportunities for subsequent insertions. By contrast, deletion attacks are restricted to existing reported edges that are controlled by compromised users under the protocol's reporting rule. This restriction is important for the upper-triangular reporting design used by the LDP triangle counting protocols considered, since each edge is reported by only one endpoint, and a compromised user can delete only the entries it is responsible for submitting.

The two directions also differ in how they use the seeding step. Seeding is meaningful for insertion attacks because a zero-gain inserted edge may create future wedges that later poisoned edges can close. However, deletion attacks do not use seeding: deleting a zero-gain edge does not create additional future triangles to remove, and it may even reduce the number of useful deletion candidates. Therefore, the negative-bias variant stops when no feasible edge has positive deletion gain.

This discussion shows that positive-bias and negative-bias attacks should not be treated as perfectly symmetric. Insertion attacks tend to be more flexible because they can exploit the large space of non-edges and the wedge-closing effect of triangle counting. Deletion attacks are more constrained because they can only remove existing reported edges controlled by compromised users. In the remainder of this section, we present protocol-specific gain derivations, primarily for the positive-bias insertion attack, which serves as the primary running example. The negative-bias variant follows the same estimator-driven principle by replacing the insertion candidate set with the deletion candidate set and defining the gain as the estimator's decrease after removing a candidate-reported edge.

\subsection{Privacy and Complexity Analysis}
\label{sec:privacy_complexity}

We briefly analyze three properties of DPA: protocol compliance, attack influence, and implementation complexity.

\myparatight{Protocol compliance}
DPA$_\triangle$  modifies report contents while preserving the protocol-specified output domain.

\begin{theorem}
\label{thm:protocol_compliance}
For any LDP triangle counting protocol with report domain $\mathcal{R}$, every poisoned report generated by DPA$_\triangle$ remains protocol-compliant, i.e., it belongs to $\mathcal{R}$.
\end{theorem}

\begin{IEEEproof}
For non-interactive protocols, users submit binary perturbed adjacency vectors or neighbor lists. DPA$_\triangle$ only changes selected entries from $0$ to $1$ for positive-bias attacks or from $1$ to $0$ for negative-bias attacks, so the report still has the same dimension, ownership constraints, and value range. For interactive protocols, the first-round structural report follows the same argument, while later-round triangle-related statistics are still submitted in the protocol-defined message format. Thus, DPA$_\triangle$ changes the content of valid reports but not their syntactic domain. Simple checks on format, dimensions, ownership, or ranges cannot identify compromised users.
\end{IEEEproof}

This theorem states protocol compliance, not that adversarial reports are sampled from the honest LDP distribution. In perturbed-report poisoning, the adversary may directly modify randomized outputs, but the resulting reports still pass validity checks. Honest users still run the prescribed local randomizer, so their $\epsilon$-edge-LDP guarantees are not weakened. 

\myparatight{Theoretical bound}
\replaced{Let $B=n_c c$ be the total poisoning budget, and let $\hat{T}$ and $\hat{T}_p$ denote the clean and poisoned estimates with other protocol randomness fixed.}{Let $B=\sum_{u\in V_c}c_u$ be the total poisoning budget. Let $\hat{T}$ and $\hat{T}_p$ be the clean and poisoned LDP triangle counting estimates, respectively.}
\deleted{If DPA$_\triangle$ performs $q\le B$ manipulations and $g_t$ is the marginal gain of the $t$-th selected manipulation, then we can get a theorem that shows as follows:}

\begin{theorem}
\label{thm:attack_bound}
\replaced{Suppose DPA$_\triangle$ performs $q\le B$ edits with signed estimator changes $g_t$. If each edit changes at most $n-2$ additive terms by at most $\Lambda$ each in absolute value, then}{Let DPA$_\triangle$ perform $q\le B$ manipulations, and let $g_t$ be the marginal gain of the $t$-th manipulation. Then the estimate shift satisfies}
\begin{equation}
\begin{aligned}
|\hat{T}_p-\hat{T}| \mathreplaced{=\left|\sum_{t=1}^{q}g_t\right|}{=\sum_{t=1}^{q}g_t}
\mathadded{\le\sum_{t=1}^{q}|g_t|}\le B(n-2)\Lambda.
\end{aligned}
\end{equation}
\deleted{For triangle counting, since one manipulated edge can affect at most $n-2$ triples, an upper bound is}
\deleted{where $\Lambda$ denotes the maximum calibration contribution of one affected triangle-related term.}
\end{theorem}

\begin{IEEEproof}
\replaced{Summing the successive changes gives $\hat{T}_p-\hat{T}=\sum_{t=1}^{q}g_t$. The triangle inequality and $|g_t|\le(n-2)\Lambda$ yield the bound.}{DPA$_\triangle$ updates the poisoned view sequentially. By definition, the selected manipulation at step $t$ changes the current calibrated estimate by $g_t$. Summing over all $q$ steps gives the first equality. If every selected gain is at most $\Gamma$, the shift is at most $q\Gamma\le B\Gamma$. Since one manipulated edge can affect at most $n-2$ node triples, and each affected triangle-related term contributes at most $\Lambda$ after calibration, the upper triangle counting bound follows.}
\end{IEEEproof}

\added{The upper bound describes worst-case sensitivity and may be loose when different edits cannot attain their maximal contributions simultaneously. Without further assumptions, the only uniform lower bound is $|\hat{T}_p-\hat{T}|\ge0$. Feasible edits may have zero gain, or their calibrated contributions may cancel. A strictly positive lower bound requires additional assumptions on the graph and available edits.}

\myparatight{Implementation complexity}
\replaced{Let $M\le n_c(n-1)$ be the number of feasible candidates and $\tau_g$ the cost of one gain evaluation. Computing initial gains costs $O(M\tau_g)$. For at most $B=n_c c$ edits, an indexed max-heap reduces candidate extraction from $O(BM)$ with linear scans to $O(B\log(M+1))$. If an edit changes or invalidates $k$ candidate entries, refreshing their gains and maintaining the heap costs $O(k(\tau_g+\log(M+1)))$. For TCountRR, $\tau_g=O(n)$. If all gains are recomputed, their $O(M\tau_g)$ cost per edit remains. Seeding retains its original search cost. Candidate and heap storage use $O(M)$ space, while dense gain or graph storage requires $O(n^2)$.}{Although Algorithm~\ref{alg:DPATC} presents a dense gain matrix for clarity, DPA$_\triangle$ can be implemented over sparse candidate sets. For each compromised user $u$, let $\mathcal{C}_u$ contain candidates that may close observed wedges or be selected by the seeding rule. The memory cost is then $O\left(\sum_{u\in V_c}|\mathcal{C}_u|\right)$ rather than $O(n^2)$. Candidate edges can be maintained in a priority queue. After inserting or deleting an edge $(u,v)$, only gains involving $u$, $v$, and their affected neighborhoods need to be updated, avoiding recomputation over all node pairs.}

\subsection{Instantiating Attack on LDP Triangle Counting Protocols}
\label{sec:template_instantiations}

We instantiate the gain-driven framework on representative LDP triangle counting protocols. Instead of designing a separate attack for each protocol, we group existing protocols into two categories according to how their estimators use reported structural information: non-interactive protocols and interactive protocols. For clarity, we present the gain derivation mainly for positive-bias insertion attacks. The negative-bias deletion variant follows the same principle by replacing the insertion candidate set with existing reported edges controlled by compromised users and defining the gain as the decrease in the corresponding estimator.

\myparatight{Non-interactive protocols}
In non-interactive protocols, each user perturbs its local structural information once and submits the perturbed report to the aggregator. The aggregator then reconstructs a noisy graph or calibrated edge statistics and estimates the triangle count through post-processing. We consider two representative protocols in this category: LocalRR$_\triangle$ and TCountRR.

For LocalRR$_\triangle$, the aggregator first reconstructs a perturbed graph and counts the four types of three-node patterns, denoted by $m_3,m_2,m_1,m_0$. The final estimate is obtained by applying the protocol-specific calibration formula to these pattern counts. Therefore, a poisoned edge affects the estimate by changing the pattern types of triples that contain this edge. Formally, if $\Psi(\cdot)$ denotes the calibrated estimator of LocalRR$_\triangle$, the marginal gain of a feasible manipulation $e$ is
\begin{equation}
\begin{aligned}
    \mathrm{Gain}(e)
    =
    \Psi(m+\Delta_e m)-\Psi(m),
\end{aligned}
\end{equation}
where $\Delta_e m$ denotes the changes in these four pattern counts caused by applying manipulation $e$.

For TCountRR, the aggregator first transforms each perturbed edge indicator into a calibrated edge variable $Y_{i,j}$ and then estimates triangles by aggregating products over node triples $\hat{T}=\sum_{i<j<k}Y_{i,j}Y_{i,k}Y_{j,k}$.
Thus, manipulating a reported edge $(u,v)$ affects all triples containing this edge. If $\Delta Y_{u,v}$ is the induced change in the calibrated edge variable, the marginal effect is
\begin{equation}
\begin{aligned}
    \mathrm{Gain}(u,v)
    =
    \Delta Y_{u,v}
    \sum_{k\in V\setminus\{u,v\}}Y_{u,k}Y_{v,k}.
\end{aligned}
\end{equation}

% For both protocols, positive-bias attacks select feasible non-existing reported edges that increase the corresponding estimator, while negative-bias attacks remove existing reported edges whose deletion decreases the estimator. The key point is that the attack is derived from the estimator used by the protocol, rather than from a separate protocol-specific heuristic.

\myparatight{Interactive protocols}
Interactive protocols, such as ARROneNS$_\triangle$, TDP-VC, and EdgeOrient$_\triangle$, improve accuracy by combining users' true local wedges with perturbed information about closing edges. Although these protocols differ in sampling, sensitivity control, and ordering constraints, their attacks follow the same principle that a poisoned edge is valuable if it changes many eligible local triangle contributions.

For a positive-bias insertion attack, the gain of inserting a perturbed edge $(u,v)$ can be written as the number of eligible local wedges closed by this edge:
\begin{equation}
\begin{aligned}
\mathrm{Gain}(u,v)
=
\sum_{w\in V\setminus\{u,v\}}
\mathbf{1}\big(\{w,u\}\in E \land \{w,v\}\in E\big),
\end{aligned}
\end{equation}
subject to the protocol-specific constraints. For example, TDP-VC restricts which sampled neighbors can contribute to the local statistic, while EdgeOrient$_\triangle$ restricts the counted wedges according to the low-out-degree ordering. For negative-bias deletion, the candidate edge must be an existing perturbed edge controlled by a compromised user, and the same local-wedge view measures how many local triangle contributions are removed.

\subsection{Partial-Knowledge Attacks}
\label{sec:partial_knowledge}

The full-knowledge setting assumes that the attacker can access the true graph structure, which may be too strong in practice. We therefore consider a \emph{partial-knowledge} attacker that only observes compromised users' local information and the perturbed information exposed by the target protocol. Following the common notion of $s$-hop neighborhoods in graph analysis~\cite{cheng2025atom,wu2022model}, we use PK(1-hop) and PK(2-hop) to describe the scope of the attacker's available knowledge.

For non-interactive protocols, the aggregator estimates triangles only from perturbed reports and never observes the true graph. Thus, the attacker can also operate on perturbed information without knowing the full true graph. In PK(1-hop), the attacker only knows the perturbed reports submitted by compromised users. In PK(2-hop), if a compromised user is adjacent to an honest user, the attacker may additionally observe the perturbed report of that honest neighbor, e.g., through observable communication channels. This extends the available perturbed information from compromised users' local reports to their 2-hop neighborhoods.

For interactive protocols, the protocol itself exposes more information because the aggregator redistributes collected perturbed structural data to users before local triangle estimation. Therefore, in PK(1-hop), the attacker can access the global perturbed graph together with the true local views of compromised users. In PK(2-hop), compromised users are further assumed to observe true connections within their 2-hop neighborhoods, as in prior local-neighborhood observation settings~\cite{wei2020asgldp,liu2024edge}. For graph regions outside the observable neighborhoods, the attacker uses the perturbed graph as a substitute for the unavailable true structure.

% Under both settings, the attacker applies the same gain-driven selection rule as in the full-knowledge attack, but computes gains using only the available partial view. Therefore, partial-knowledge attacks evaluate the robustness of LDP triangle counting protocols under a more realistic adversary that cannot fully reconstruct the true graph.

% Although the attacker cannot access the true graph structure under the assumption of partial knowledge, the perturbed information available to the attacker is identical to that available to the aggregator. As a result, the attacker can simulate the aggregator’s execution and ensure that the poisoning still achieves the intended effect on the aggregator’s estimation.

\section{Evaluation}
\subsection{Experimental Settings}
\myparatight{Datasets}
We evaluate DPA$_{\triangle}$ on four real-world graphs from the Stanford Network Analysis Project (SNAP)~\cite{snapnets}, which are widely used in triangle counting evaluation.
The datasets cover diverse network types: a social network (ego-Facebook), a collaboration network (ca-AstroPh), a citation network (cit-HepTh) and a communication network(email-Enron). Table~\ref{tab:datasets} summarizes the basic statistics of each graph. We preprocess all four datasets into undirected graphs.
% \redtext{[Optional: Specify preprocessing for reproducibility: e.g., whether graphs are treated as undirected, and whether self-loops/multi-edges are removed and/or the largest connected component is used.]}

\begin{table}[htbp]
  \caption{Details for graph datasets}
  \label{tab:datasets}
  \small
  \begin{tabular}{ccccl}
    \toprule
    Name &Type &$\lvert V\rvert$ &$\lvert E\rvert$ &$d_{max}$\\
    \midrule
    ego-Facebook &Social &4,039 &88,234 &1,045\\
    ca-AstroPh &Collaboration  &18,772 &198,110 &504 \\
    cit-HepTh &Citation &27,770 &352,285 &2,468\\
    email-Enron &Communication &36,692 &183,831 &1,383\\
    % com-Orkut &Social &3072441 &117185083 &-\\
  \bottomrule
\end{tabular}
\end{table}

\begin{table}
  \caption{Descriptions of parameter settings}
  \centering
  \label{tab:parameters}
  \small
  \begin{tabular}{ccl}
    \toprule
    Parameter &Default Value &Meaning\\
    \midrule
    $\epsilon$ &2 &Privacy Budget\\
    $c$    &20    &Attack Capability\\
    $\beta$    &0.05    &The fraction of Compromised Users\\
    % $\gamma$    &0.05    &The fraction of Target Users\\
  \bottomrule
\end{tabular}
\end{table}

\myparatight{Parameters}
The effectiveness of the attack is mainly affected by the privacy budget $\epsilon$, the per-user poisoning capability $c$, and the fraction of compromised users $\beta$. \reftab\ref{tab:parameters} lists the default values of these parameters. Those defaults apply to all experiments unless otherwise stated. In the following, we analyze the effect of each parameter by varying it while keeping the others fixed.

\myparatight{Baselines}
We compare our method against the randomized baseline attacks introduced in Section~\ref{sec:baselines}. Specifically, RPA$_{\triangle}$ randomly changes $c$ edges that don’t exist in the perturbed neighbor list of each compromised user to exist. In contrast, RTA$_{\triangle}$ first randomly changes $c$ edges that don’t exist in the true neighbor list of each compromised user, and then applies the LDP mechanism to the poisoned true neighbor list before submission.

\begin{figure}[ht]
\tabcolsep=0cm
\setlength{\abovecaptionskip}{0.2cm}
\setlength{\belowcaptionskip}{-0.2cm}
\centering
\begin{tabular}{cc}
\multicolumn{2}{c}{\includegraphics[width=0.4\textwidth]{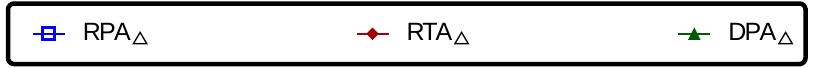}}\\
\includegraphics[width=0.25\textwidth]{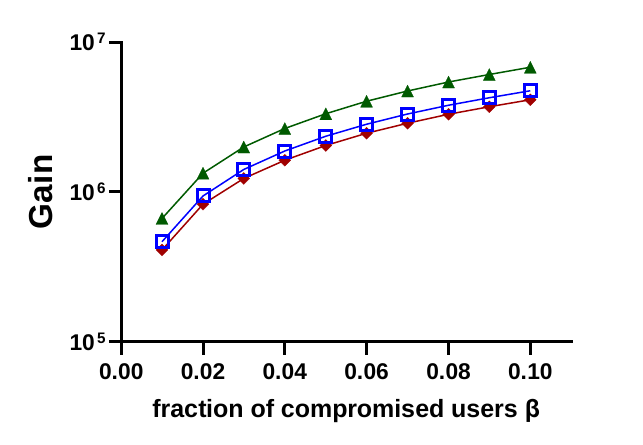}
& \includegraphics[width=0.25\textwidth]{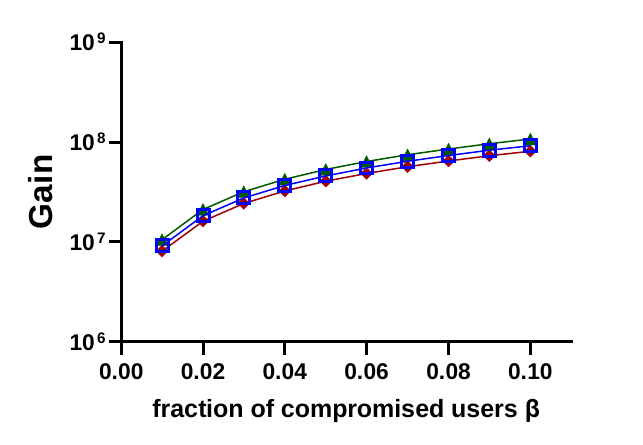}\\
     \graphSize{(a) ego-Facebook}
&    \graphSize{(b) ca-AstroPh}\\
 \includegraphics[width=0.25\textwidth]{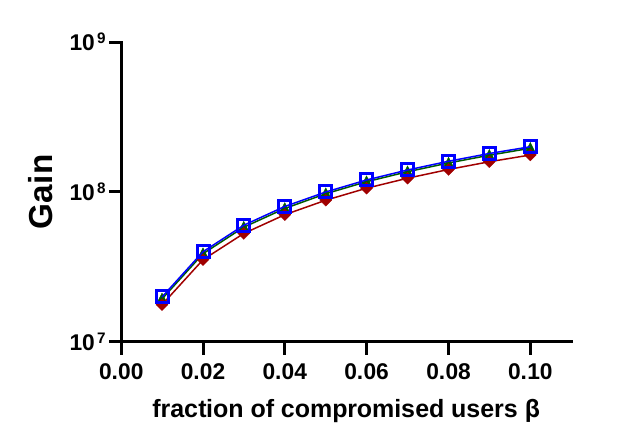}  
& \includegraphics[width=0.25\textwidth]{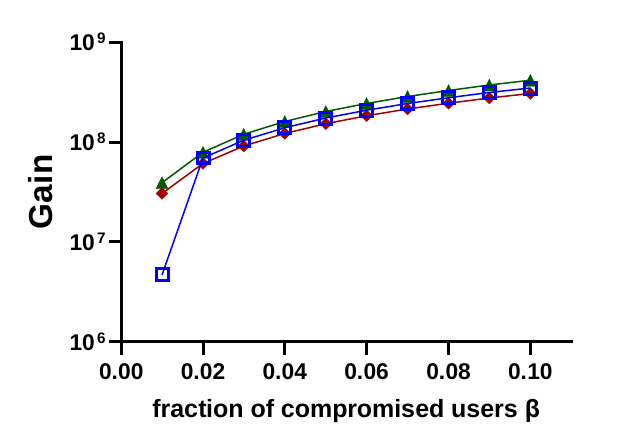}  \\
   \graphSize{(c) cit-HepTh}
&    \graphSize{(d) email-Enron}\\
\end{tabular}
\caption{Attacking LocalRR$_\triangle$}
\label{fig:LocalRR_ff}
\end{figure}

\subsection{Results for Non-Interactive Protocols}

\begin{figure}[ht]
\tabcolsep=0cm
\setlength{\abovecaptionskip}{0.2cm}
\setlength{\belowcaptionskip}{-0.2cm}
\centering
\begin{tabular}{cc}
\multicolumn{2}{c}{\includegraphics[width=0.4\textwidth]{experimentFigures/legend1.pdf}}\\
\includegraphics[width=0.25\textwidth]{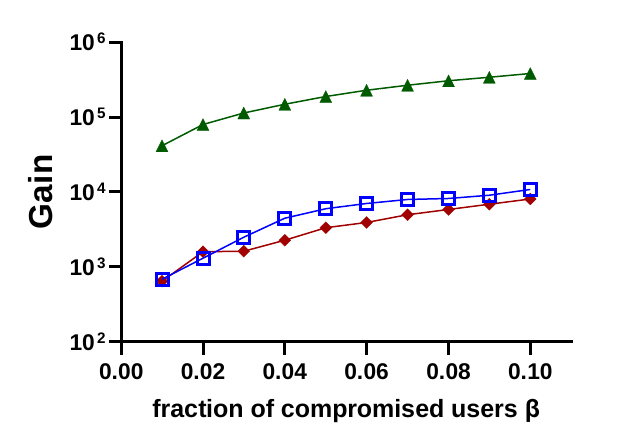}
& \includegraphics[width=0.25\textwidth]{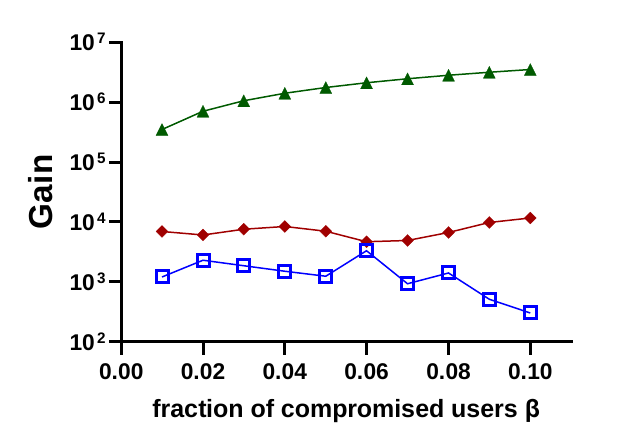}\\
     \graphSize{(a) ego-Facebook}
&    \graphSize{(b) ca-AstroPh}\\
 \includegraphics[width=0.25\textwidth]{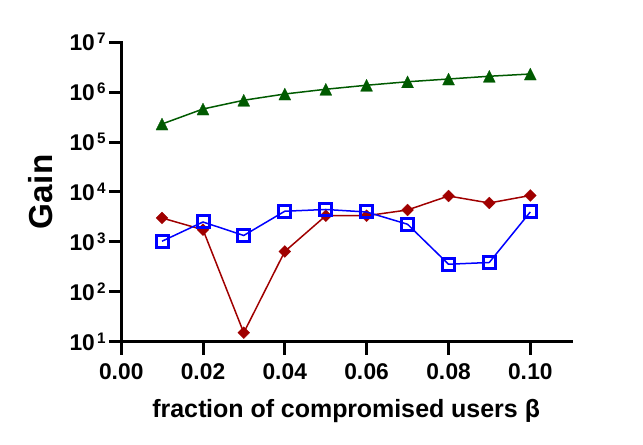}  
& \includegraphics[width=0.25\textwidth]{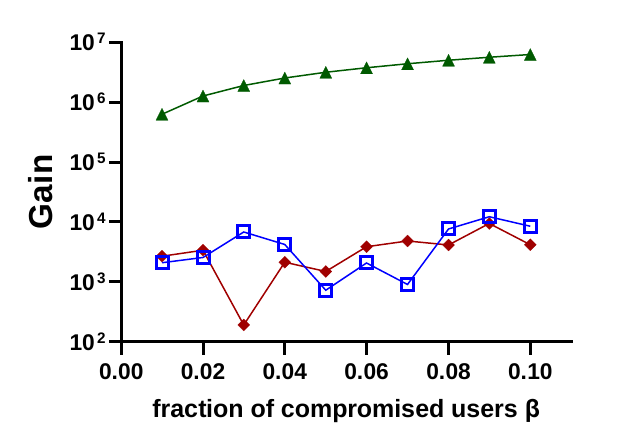}  \\
   \graphSize{(c) cit-HepTh}
&    \graphSize{(d) email-Enron}\\
\end{tabular}
\caption{Attacking TCountRR}
\label{fig:TCountRR_ff}
\end{figure}

We first evaluate the performance of non-interactive LDP triangle counting protocols under data poisoning attacks with different fractions of compromised users. As shown in \reffig\ref{fig:LocalRR_ff}, all attack methods significantly affect LocalRR$_{\triangle}$, indicating that LocalRR$_{\triangle}$ is highly vulnerable to data poisoning. \reffig\ref{fig:TCountRR_ff} further shows that TCountRR is relatively robust against randomized attacks, but remains vulnerable to proposed attack strategies. 
Meanwhile, the attack gains achieved by the two randomized attacks fluctuate substantially, whereas DPA$_{\triangle}$ consistently attains higher attack gains as the fraction of compromised users increases. This further demonstrates the effectiveness of DPA$_{\triangle}$.
The main reason is that, in non-interactive protocols, the aggregator can perform estimation only based on perturbed data. In this situation, there is no effective method to distinguish poisoned data from perturbed data or to mitigate their impact once they are mixed together.

\subsection{Results for Interactive Protocols}

\begin{figure}[ht]
\tabcolsep=0cm
\setlength{\abovecaptionskip}{0.2cm}
\setlength{\belowcaptionskip}{-0.2cm}
\centering
\begin{tabular}{cc}
\multicolumn{2}{c}{\includegraphics[width=0.4\textwidth]{experimentFigures/legend1.pdf}}\\
\includegraphics[width=0.25\textwidth]{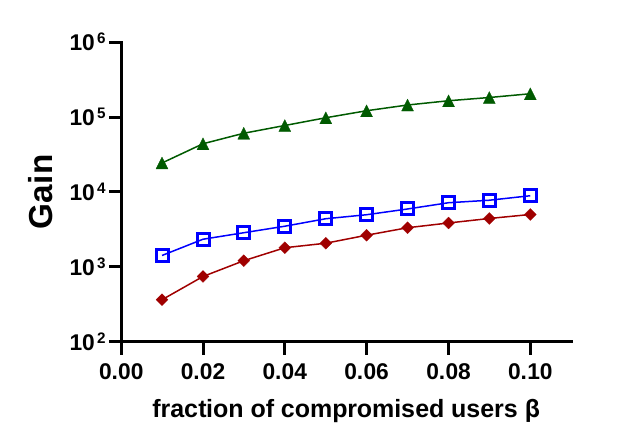}
& \includegraphics[width=0.25\textwidth]{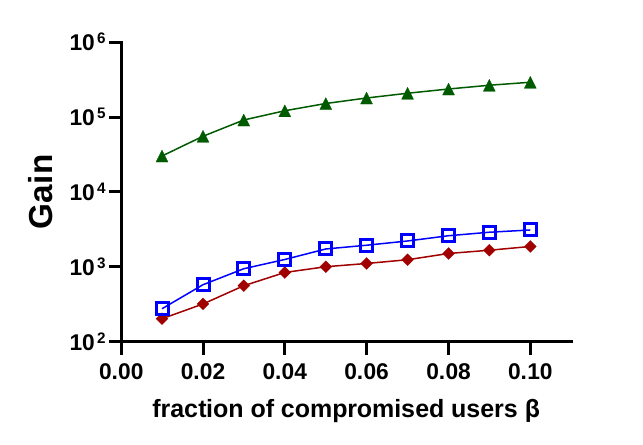}\\
     \graphSize{(a) ego-Facebook}
&    \graphSize{(b) ca-AstroPh}\\
 \includegraphics[width=0.25\textwidth]{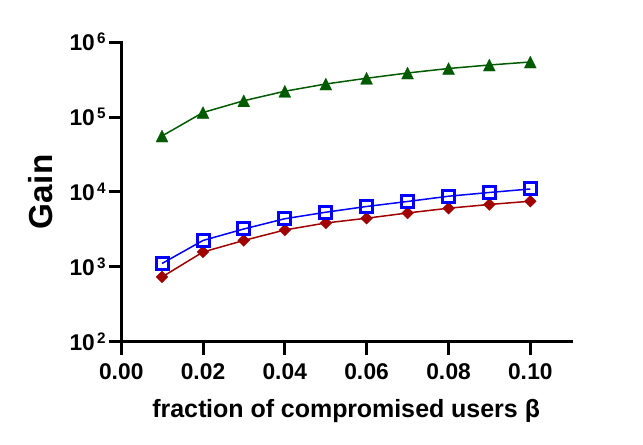}  
& \includegraphics[width=0.25\textwidth]{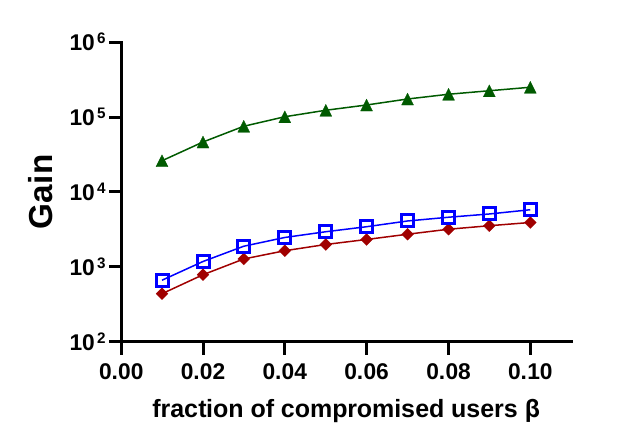}  \\
   \graphSize{(c) cit-HepTh}
&    \graphSize{(d) email-Enron}\\
\end{tabular}
\caption{Attacking ARROneNS$_\triangle$}
\label{fig:ARROneNS_ff}
\end{figure}

\begin{figure}[ht]
\tabcolsep=0cm
\setlength{\abovecaptionskip}{0.2cm}
\setlength{\belowcaptionskip}{-0.2cm}
\centering
\begin{tabular}{cc}
\multicolumn{2}{c}{\includegraphics[width=0.4\textwidth]{experimentFigures/legend1.pdf}}\\
\includegraphics[width=0.25\textwidth]{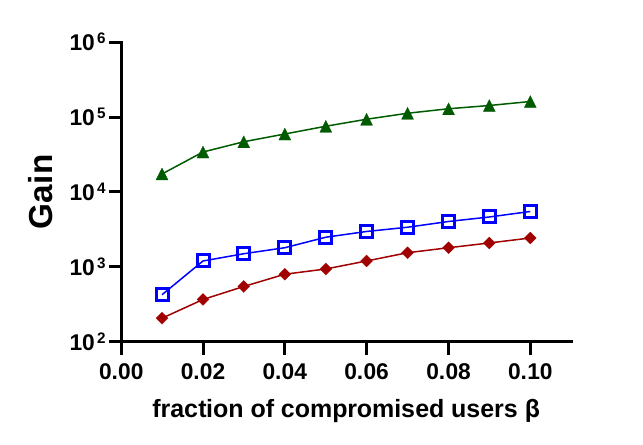}
& \includegraphics[width=0.25\textwidth]{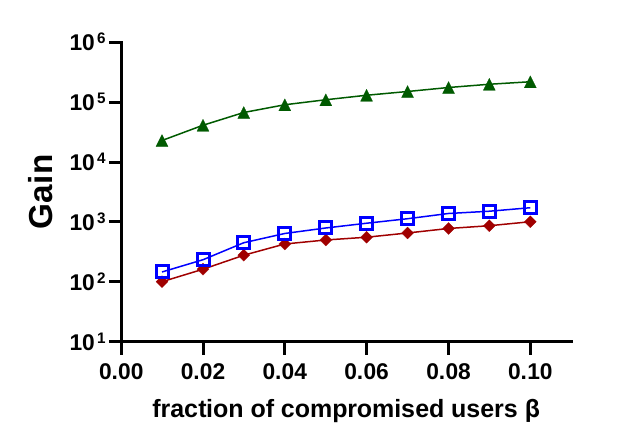}\\
     \graphSize{(a) ego-Facebook}
&    \graphSize{(b) ca-AstroPh}\\
 \includegraphics[width=0.25\textwidth]{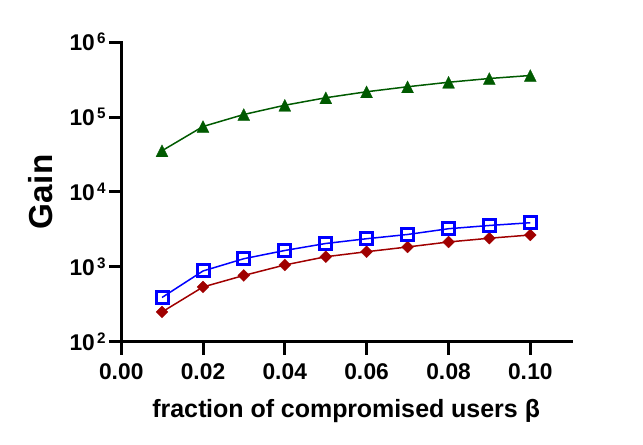}  
& \includegraphics[width=0.25\textwidth]{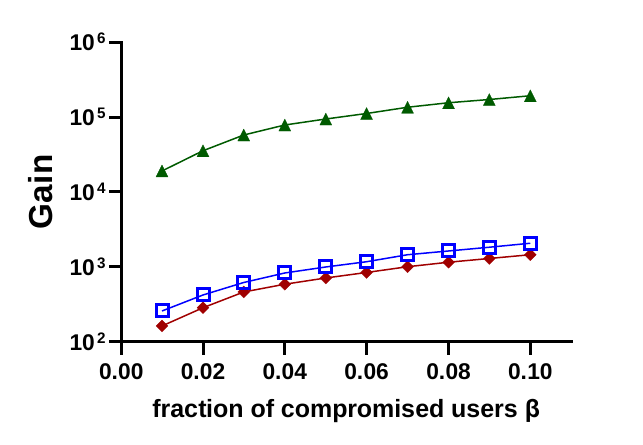}  \\
   \graphSize{(c) cit-HepTh}
&    \graphSize{(d) email-Enron}\\
\end{tabular}
\caption{Attacking TDP-VC}
\label{fig:TDPVC_ff}
\end{figure}

\begin{figure}[ht]
\tabcolsep=0cm
\setlength{\abovecaptionskip}{0.2cm}
\setlength{\belowcaptionskip}{-0.2cm}
\centering
\begin{tabular}{cc}
\multicolumn{2}{c}{\includegraphics[width=0.4\textwidth]{experimentFigures/legend1.pdf}}\\
\includegraphics[width=0.25\textwidth]{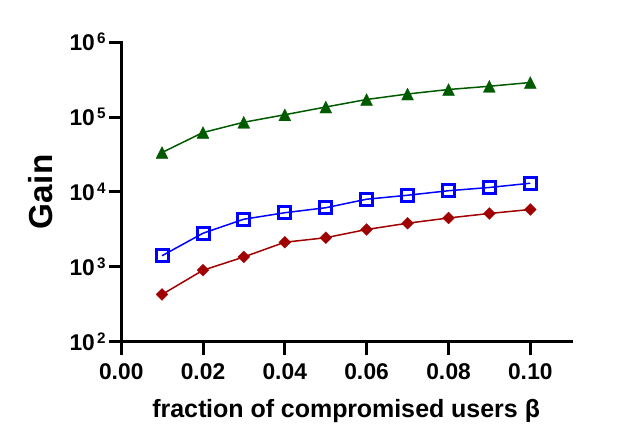}
& \includegraphics[width=0.25\textwidth]{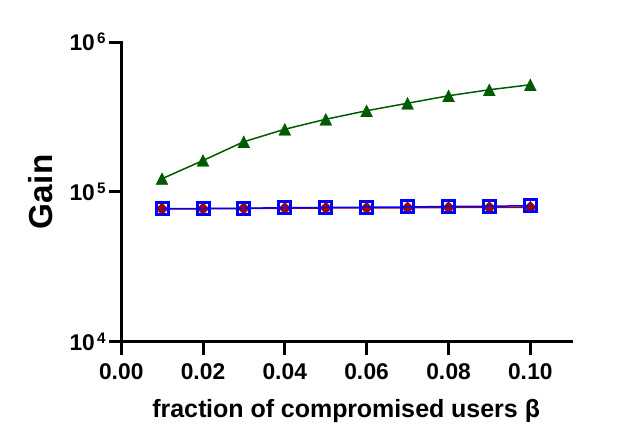}\\
     \graphSize{(a) ego-Facebook}
&    \graphSize{(b) ca-AstroPh}\\
 \includegraphics[width=0.25\textwidth]{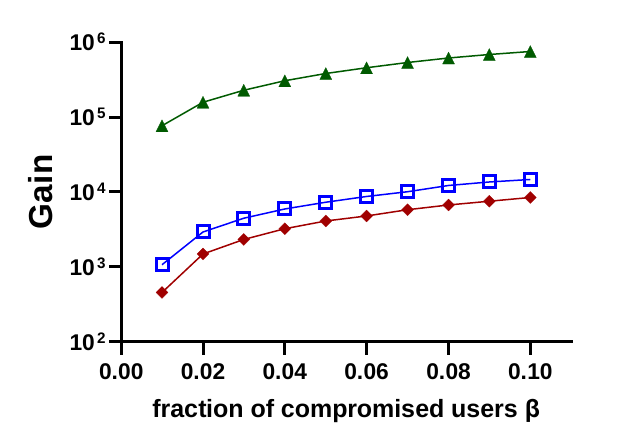}  
& \includegraphics[width=0.25\textwidth]{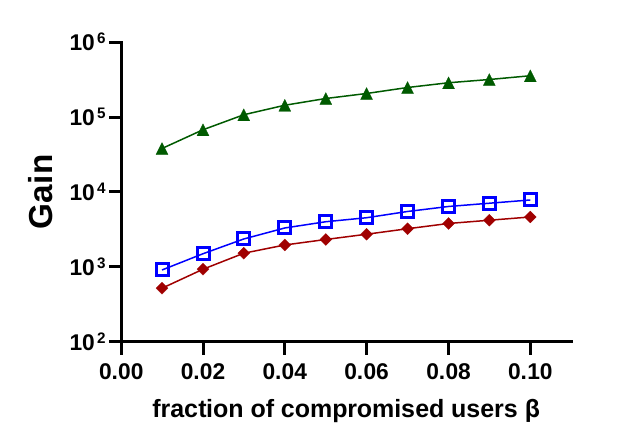}  \\
   \graphSize{(c) cit-HepTh}
&    \graphSize{(d) email-Enron}\\
\end{tabular}
\caption{Attacking EdgeOrient$_\triangle$}
\label{fig:EdgeOrient_ff}
\end{figure}

The attack performance of different strategies against interactive LDP triangle counting protocols under varying fractions of compromised users is shown in \reffig\ref{fig:ARROneNS_ff}-\ref{fig:EdgeOrient_ff}. The results demonstrate that the proposed attack strategy, DPA$_{\triangle}$, remains effective against interactive protocols. In addition, the impact of the two randomized attacks on interactive protocols is relatively limited, and the attack gain achieved by DPA$_{\triangle}$ is also smaller than that in the non-interactive setting. This is because interactive protocols exploit the local views of honest users, thereby mitigating the effect of poisoned data.

\subsection{Impact of Different Parameters}

\begin{figure}[ht]
\tabcolsep=0cm
\setlength{\abovecaptionskip}{0.2cm}
\setlength{\belowcaptionskip}{-0.2cm}
\centering
\begin{tabular}{cc}
\multicolumn{2}{c}{\includegraphics[width=0.4\textwidth]{experimentFigures/legend1.pdf}}\\
\includegraphics[width=0.25\textwidth]{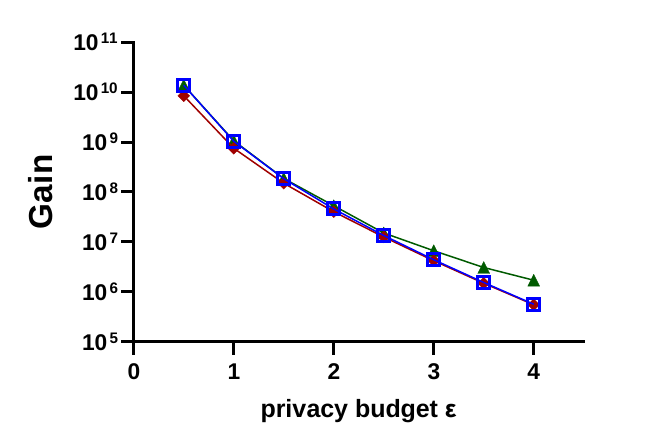}
& \includegraphics[width=0.25\textwidth]{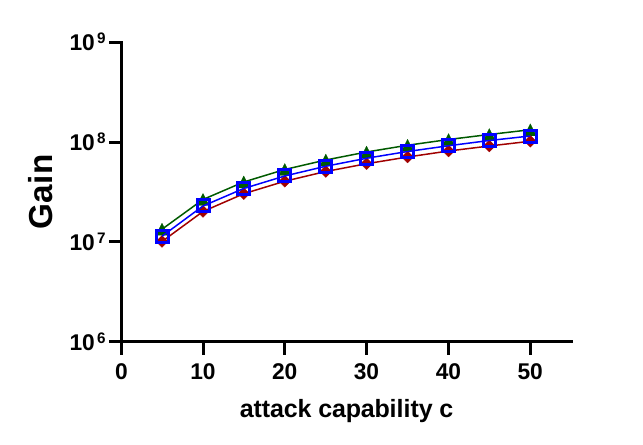} \\
     \graphSize{(a) LocalRR$_{\triangle}$,$\epsilon$}
&    \graphSize{(b) LocalRR$_{\triangle}$,$c$}\\
\includegraphics[width=0.25\textwidth]{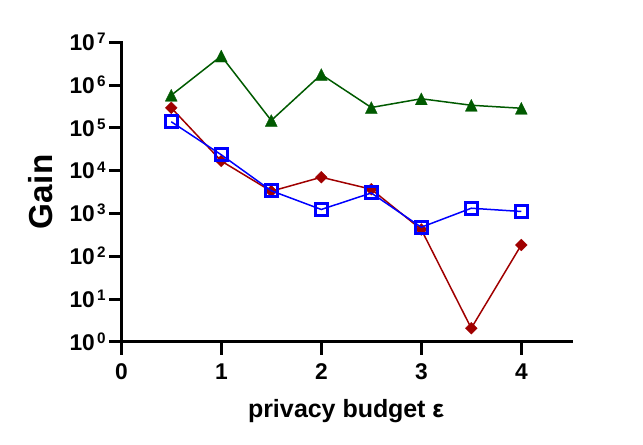}
& \includegraphics[width=0.25\textwidth]{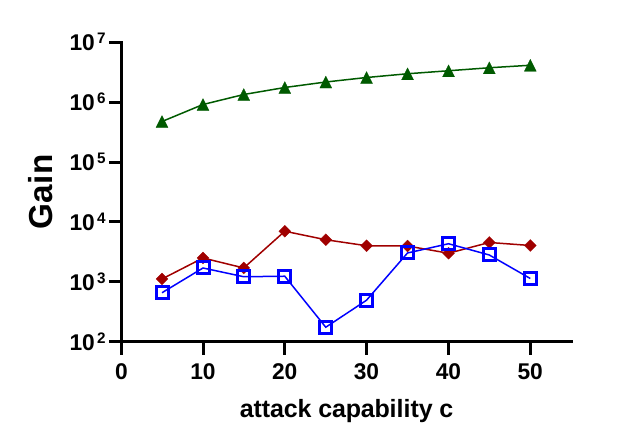} \\
     \graphSize{(c) TCountRR,$\epsilon$}
&    \graphSize{(d) TCountRR,$c$}\\
\end{tabular}
\caption{Performance of attacking strategies for LDP non-interactive triangle counting protocols on ca-AstroPh with changing $\epsilon$ and $c$}
\label{fig:parameters_nint}
\end{figure}

\begin{figure}[ht]
\tabcolsep=0cm
\setlength{\abovecaptionskip}{0.2cm}
\setlength{\belowcaptionskip}{-0.2cm}
\centering
\begin{tabular}{cc}
\multicolumn{2}{c}{\includegraphics[width=0.4\textwidth]{experimentFigures/legend1.pdf}}\\
\includegraphics[width=0.25\textwidth]{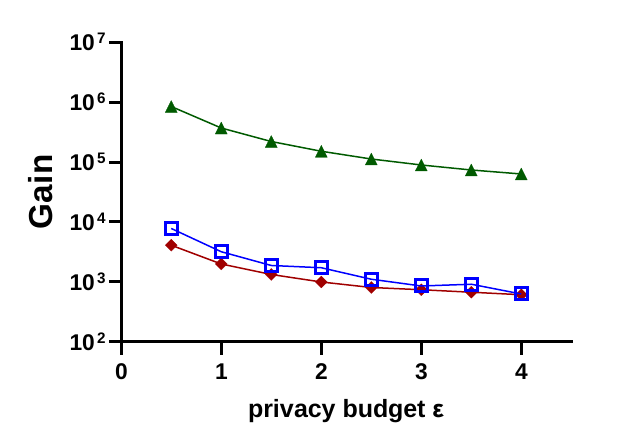}
& \includegraphics[width=0.25\textwidth]{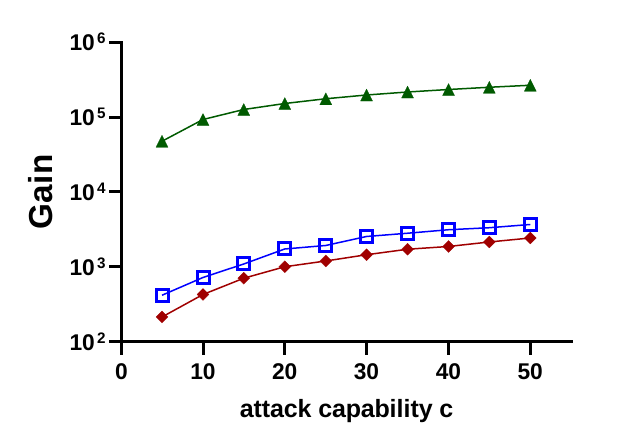} \\
     \graphSize{(a) ARROneNS$_\triangle$, $\epsilon$}
&    \graphSize{(b) ARROneNS$_\triangle$, $c$}\\
\includegraphics[width=0.25\textwidth]{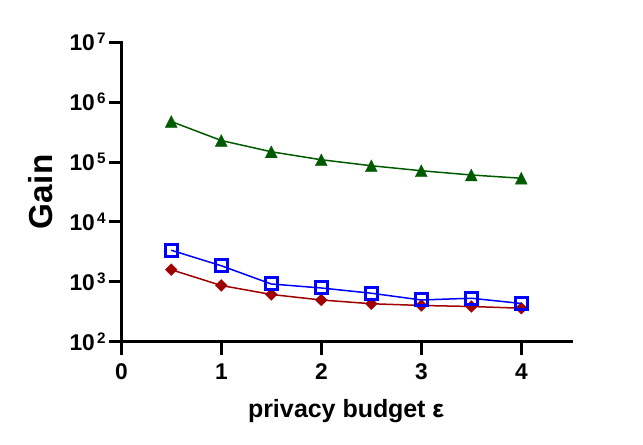}
& \includegraphics[width=0.25\textwidth]{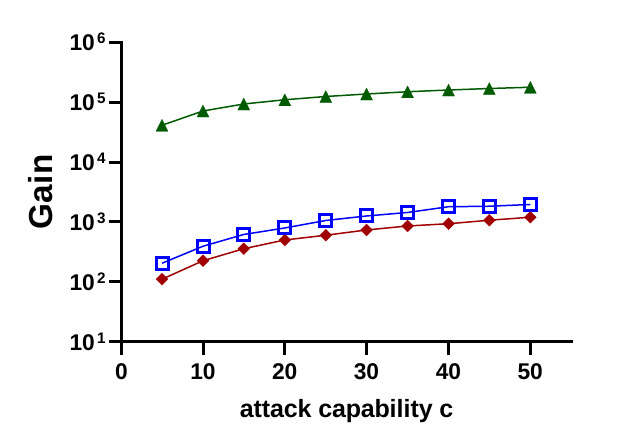} \\
    \graphSize{(c) TDP-VC, $\epsilon$}
&    \graphSize{(d) TDP-VC, $c$}\\
\includegraphics[width=0.25\textwidth]{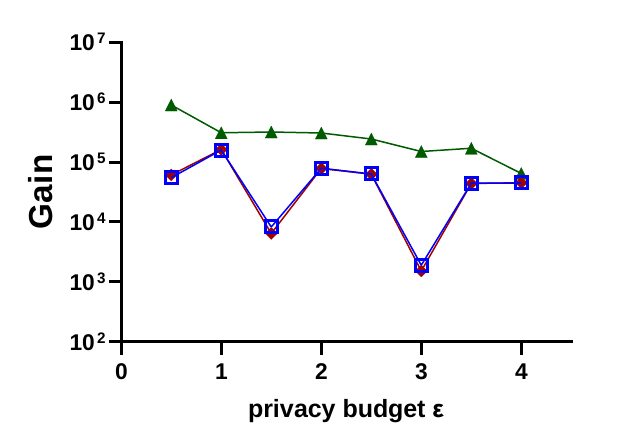}
& \includegraphics[width=0.25\textwidth]{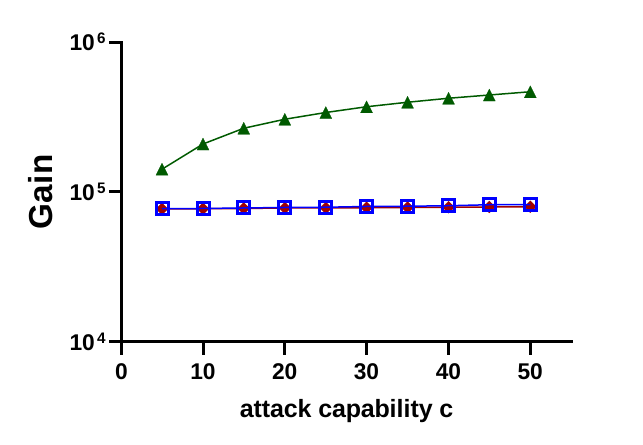} \\
    \graphSize{(e) EdgeOrient$_\triangle$, $\epsilon$}
&    \graphSize{(f) EdgeOrient$_\triangle$, $c$}\\
\end{tabular}
\caption{Performance of attacking strategies for LDP interactive triangle counting protocols on ca-AstroPh with changing $\epsilon$ and $c$}
\label{fig:parameters_int}
\end{figure}

We next examine how different parameter settings affect the performance of the proposed attacks on triangle counting. Specifically, we vary the privacy budget $\epsilon$ from 0.5 to 4 and the attack capability of each compromised user from 5 to 50, and evaluate the resulting attack effectiveness. \reffig\ref{fig:parameters_nint} shows the performance of non-interactive LDP triangle counting protocols on ca-AstroPh under the three attack strategies. For LocalRR$_{\triangle}$, we observe that the attack gain decreases as the privacy budget $\epsilon$ increases. This suggests that under a smaller privacy budget, poisoned data are more concealed and can induce a larger attack gain against LocalRR$_{\triangle}$. In addition, as the attack capability $c$ increases, the attack gain on LocalRR$_{\triangle}$ also increases. For TCountRR, although the impact of DPA$_{\triangle}$ exhibits minor fluctuations as $\epsilon$ increases, the overall trend is still downward. As $c$ increases, the effects of the two randomized attacks fluctuate, whereas the attack gain achieved by DPA$_{\triangle}$ increases steadily and remains substantially higher than those of the two randomized baselines.

\reffig\ref{fig:parameters_int} presents the performance of the three interactive LDP triangle counting protocols under data poisoning attacks with different privacy budgets and attack capabilities. For ARROneNS$_{\triangle}$ and TDP-VC, the attack gains of all three strategies decrease as the privacy budget increases and increase as the attack capability grows. For EdgeOrient$_{\triangle}$, although the gains of the two randomized attacks fluctuate with the privacy budget, the attack gain of DPA$_{\triangle}$ decreases steadily as the privacy budget increases. Moreover, the two randomized attacks show no clear improvement as the attack capability increases. This is because the low-out-degree ordering in EdgeOrient$_{\triangle}$ restricts each user to counting triangles associated with only a subset of its neighbors rather than all of them. This fundamentally limits the number of triangles that each user can contribute. As a result, even if the attack capability of compromised users increases, the gains of randomized attacks do not improve significantly. In contrast, DPA$_{\triangle}$ prioritizes the neighbors that yield the highest gain when selecting local triangles to count. Therefore, its attack gain increases steadily as the attack capability grows.
% Overall, the effectiveness of data poisoning attacks against interactive LDP triangle counting protocols decreases as the privacy budget increases and increases as the attack capability grows.

\begin{figure*}[ht]
\tabcolsep=0cm
\setlength{\abovecaptionskip}{0.2cm}
\setlength{\belowcaptionskip}{-0.2cm}
\centering
\begin{tabular}{ccccc}
\multicolumn{5}{c}{\includegraphics[width=0.4\textwidth]{experimentFigures/legend1.pdf}}\\
\includegraphics[width=0.2\textwidth]{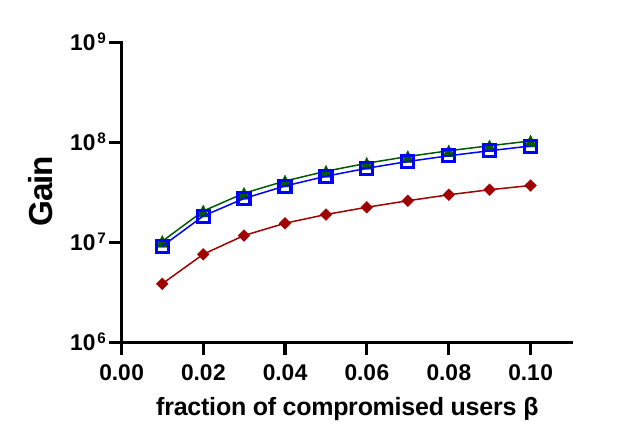} &\includegraphics[width=0.2\textwidth]{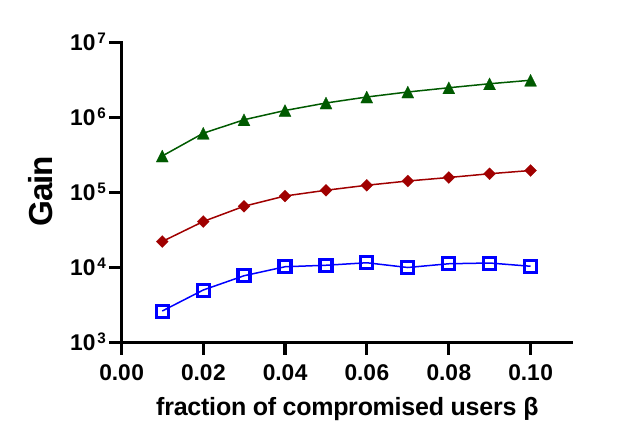}
&\includegraphics[width=0.2\textwidth]{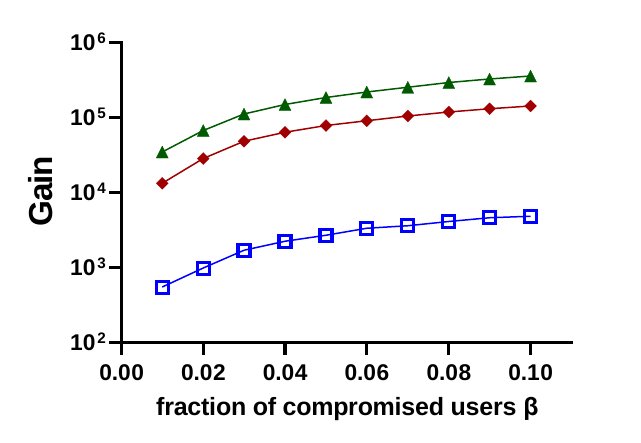}
&\includegraphics[width=0.2\textwidth]{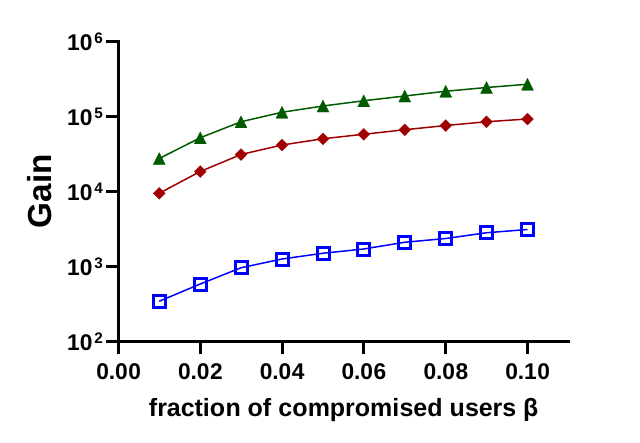}
&\includegraphics[width=0.2\textwidth]{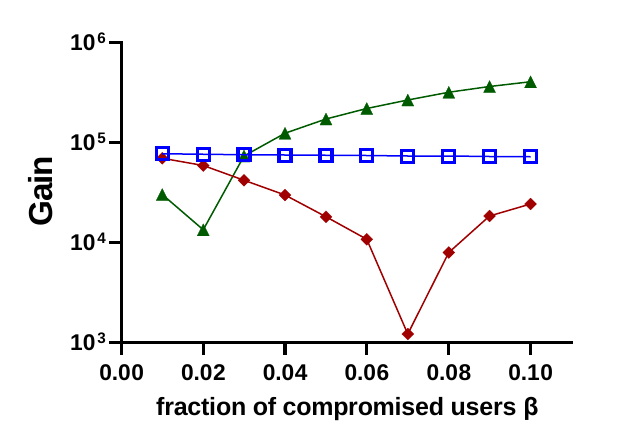}\\
\graphSize{(a) LocalRR$_\triangle$, $\beta$}
&\graphSize{(b) TCountRR, $\beta$}
&\graphSize{(c) ARROneNS$_\triangle$, $\beta$}
&\graphSize{(d) TDP-VC, $\beta$}
&\graphSize{(e) EdgeOrient$_\triangle$, $\beta$}\\
\includegraphics[width=0.2\textwidth]{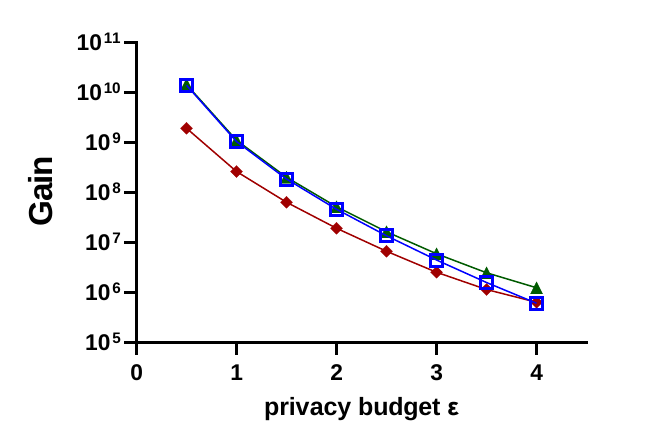} &\includegraphics[width=0.2\textwidth]{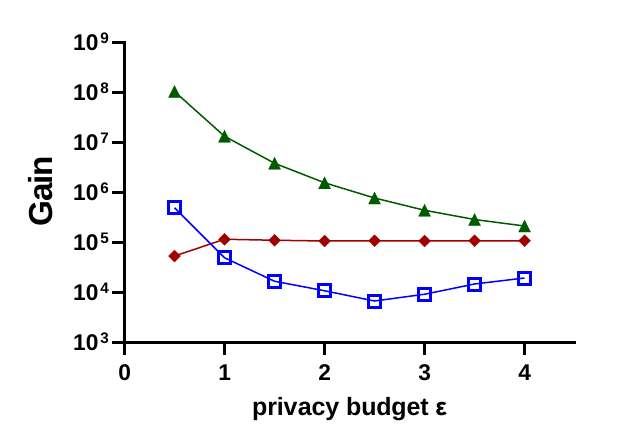}
&\includegraphics[width=0.2\textwidth]{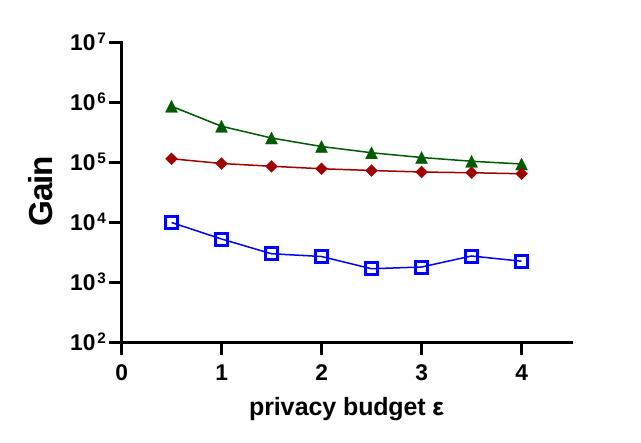}
&\includegraphics[width=0.2\textwidth]{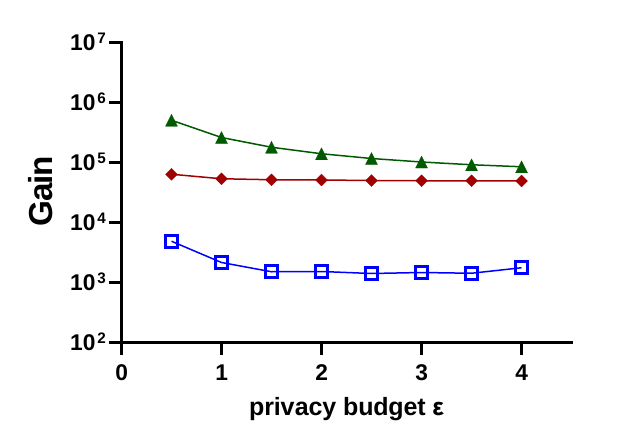}
&\includegraphics[width=0.2\textwidth]{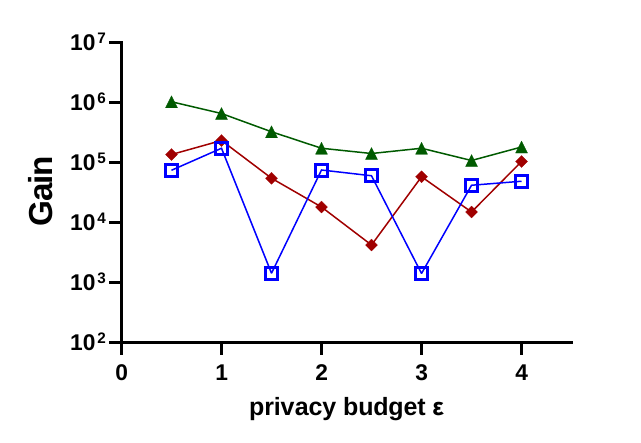}\\
\graphSize{(f) LocalRR$_\triangle$, $\epsilon$}
&\graphSize{(g) TCountRR, $\epsilon$}
&\graphSize{(h) ARROneNS$_\triangle$, $\epsilon$}
&\graphSize{(i) TDP-VC, $\epsilon$}
&\graphSize{(j) EdgeOrient$_\triangle$, $\epsilon$}\\
\includegraphics[width=0.2\textwidth]{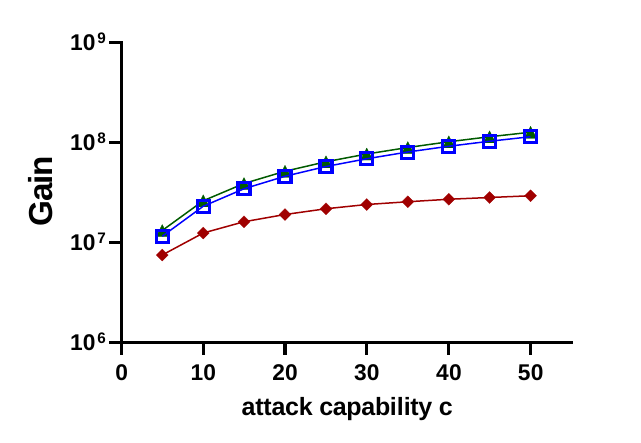} &\includegraphics[width=0.2\textwidth]{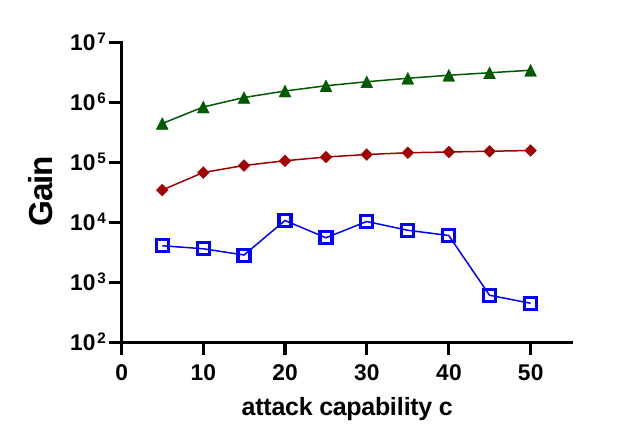}
&\includegraphics[width=0.2\textwidth]{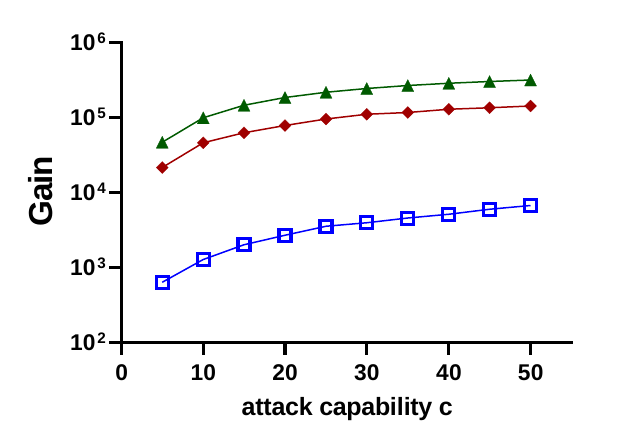}
&\includegraphics[width=0.2\textwidth]{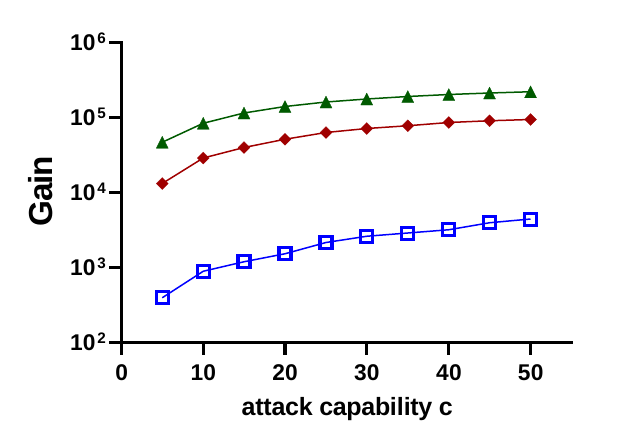}
&\includegraphics[width=0.2\textwidth]{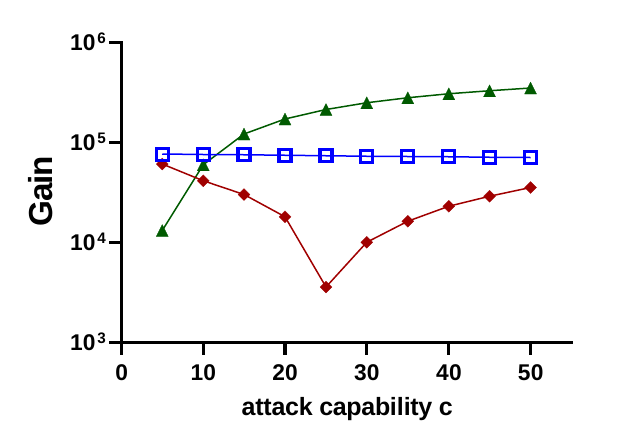}\\
\graphSize{(k) LocalRR$_\triangle$, $c$}
&\graphSize{(l) TCountRR, $c$}
&\graphSize{(m) ARROneNS$_\triangle$, $c$}
&\graphSize{(n) TDP-VC, $c$}
&\graphSize{(o) EdgeOrient$_\triangle$, $c$}\\
\end{tabular}
\caption{Performance of negative-bias data poisoning attacks on ca-AstroPh}
\label{fig:neg_attack}
\end{figure*}

\begin{figure*}[ht]
\tabcolsep=0cm
\setlength{\abovecaptionskip}{0.2cm}
\setlength{\belowcaptionskip}{-0.2cm}
\centering
\begin{tabular}{ccccc}
\multicolumn{5}{c}{\includegraphics[width=0.4\textwidth]{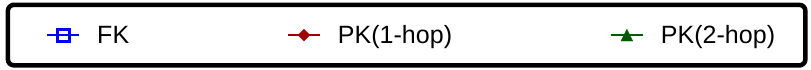}}\\
\includegraphics[width=0.2\textwidth]{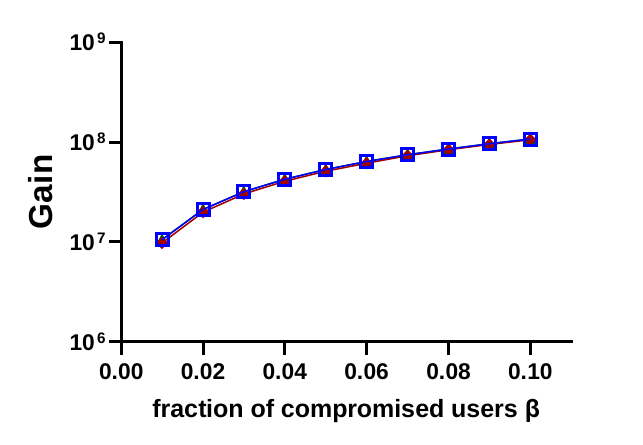}
& \includegraphics[width=0.2\textwidth]{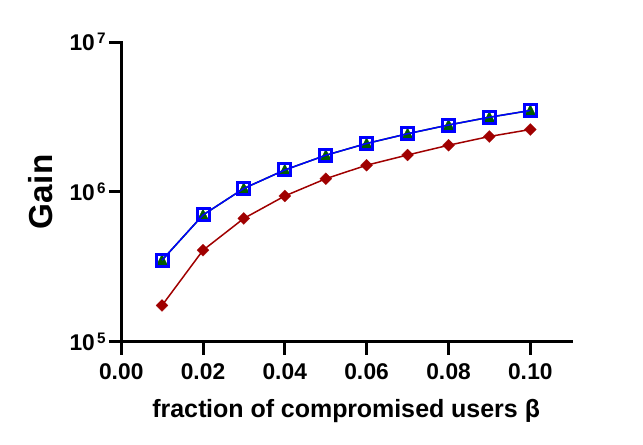}
& \includegraphics[width=0.2\textwidth]{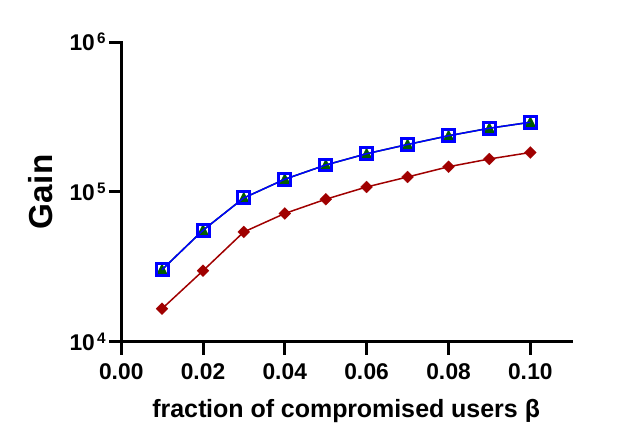}
& \includegraphics[width=0.2\textwidth]{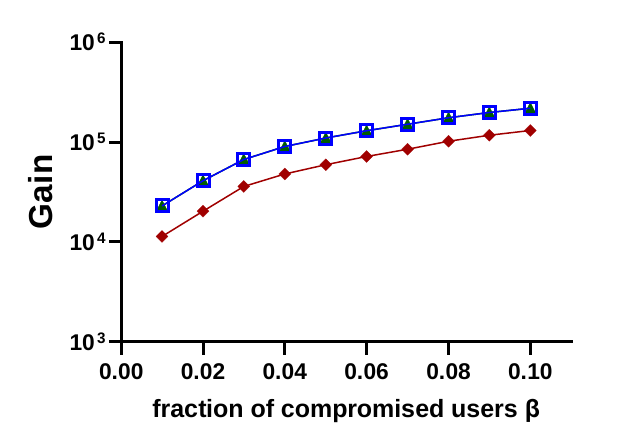}
& \includegraphics[width=0.2\textwidth]{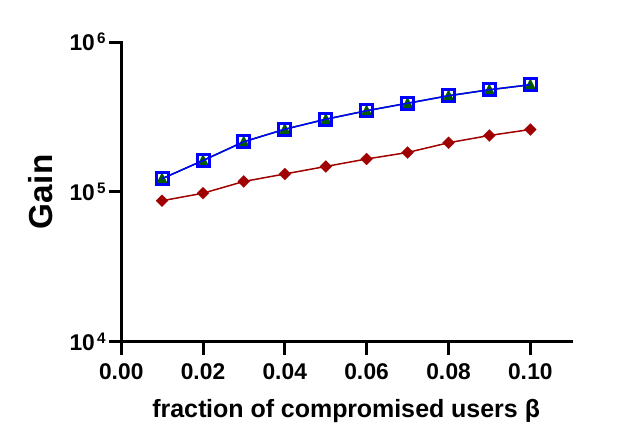}
\\
\graphSize{(a) LocalRR$_\triangle$, $\beta$}
& \graphSize{(b) TCountRR, $\beta$}
& \graphSize{(c) ARROneNS$_\triangle$, $\beta$}
& \graphSize{(d) TDP-VC, $\beta$}
& \graphSize{(e) EdgeOrient$_\triangle$, $\beta$}
\\
% \multicolumn{2}{c}{$k$-star}
\end{tabular}
\caption{Attacking LDP triangle counting protocols with partial knowledge on ca-AstroPh with changing $\beta$}
\label{fig:partial_triangle}
\end{figure*}

\subsection{Negative-Bias Data Poisoning Attacks}
\label{sec:eval_negative_bias}

We further evaluate DPA$_\triangle$ under the negative-bias setting, where the attack goal is to decrease the final triangle counting estimate by deleting edges. We compare three deletion-based poisoning attacks: RPA$_\triangle$, RTA$_\triangle$, and DPA$_\triangle$. RPA$_\triangle$ randomly deletes existing edges from perturbed reports, RTA$_\triangle$ randomly deletes edges from compromised users' true local views before randomization, and DPA$_\triangle$ greedily deletes reported edges with the largest marginal decrease in the calibrated estimator.

Fig.~\ref{fig:neg_attack} reports the attack gain under different compromised-user fractions, privacy budgets, and attack capabilities. Overall, DPA$_\triangle$ achieves the largest gain in most settings, especially on TCountRR, ARROneNS$_\triangle$, and TDP-VC. This shows that estimator-aware edge deletion is more effective than random deletion, even when all attacks are restricted to the negative-bias setting. As the compromised-user fraction $\beta$ or the attack capability $c$ increases, the gain of DPA$_\triangle$ generally increases because the attacker has more opportunities to remove structurally important reported edges.

The effect of the privacy budget $\epsilon$ is more protocol-dependent, since $\epsilon$ changes both the noise level and the calibration sensitivity of the estimator. The random baselines exhibit larger fluctuations, particularly on LocalRR$_\triangle$ and EdgeOrient$_\triangle$, because deletion attacks can only operate on existing candidate edges and are therefore sensitive to the available reported structure. These results confirm that the selection of deleted edges substantially affects the effectiveness of negative-bias poisoning and that the proposed strategy yields a more reliable deletion attack than random baselines.

\subsection{Evaluate Attacks with Partial Knowledge}
\label{sec:eval_partial_knowledge}

We further evaluate DPA$_\triangle$ under partial-knowledge settings with different fractions of compromised users. We consider three knowledge levels. FK assumes that the attacker has full knowledge required by the attack. PK(1-hop) only uses information available from compromised users' local views or reports. PK(2-hop) additionally includes information within compromised users' 2-hop neighborhoods. For unavailable regions, the attacker uses the perturbed graph information exposed by the protocol.

\reffig~\ref{fig:partial_triangle}(a)(b) show the results on non-interactive protocols. DPA$_\triangle$ still achieves considerable gain under PK(1-hop), indicating that compromised users' own perturbed reports already provide useful structural signals. When the knowledge expands to PK(2-hop), the performance approaches FK because more triangle-related structures around compromised users become observable.

\reffig~\ref{fig:partial_triangle}(c)(d)(e) show the results on interactive protocols. DPA$_\triangle$ remains effective under both PK(1-hop) and PK(2-hop), and the gain generally increases with the fraction of compromised users. This is because interactive protocols expose perturbed structural information through the interaction, allowing the attacker to approximate gain-driven selection without full access to the true graph.

\begin{figure*}[ht]
\tabcolsep=0cm
\setlength{\abovecaptionskip}{0.2cm}
\setlength{\belowcaptionskip}{-0.2cm}
\centering
\begin{tabular}{ccccc}
\multicolumn{5}{c}{\includegraphics[width=0.4\textwidth]{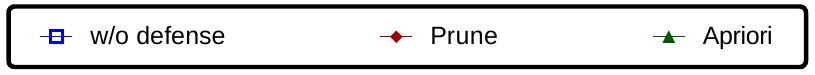}}\\
\includegraphics[width=0.2\textwidth]{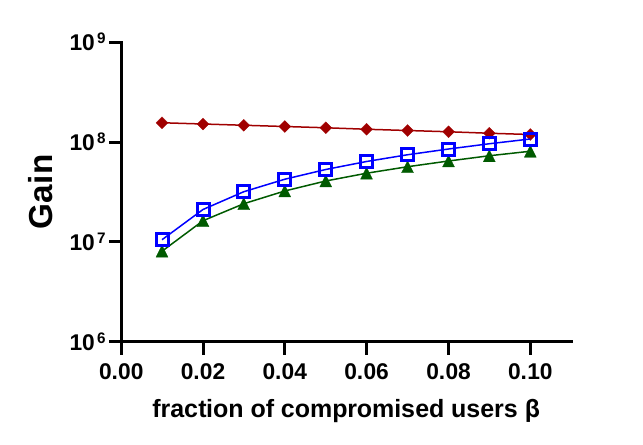}
& \includegraphics[width=0.2\textwidth]{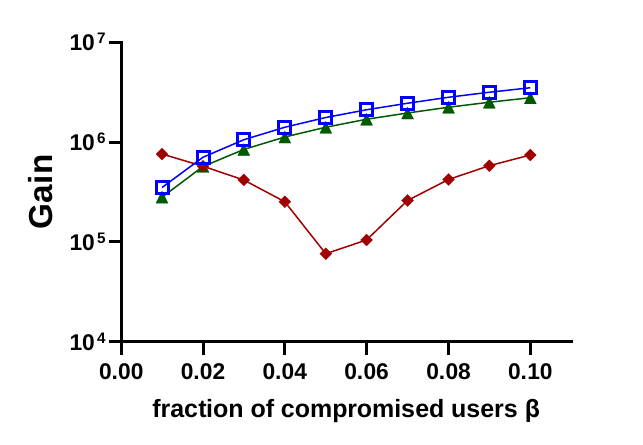}
& \includegraphics[width=0.2\textwidth]{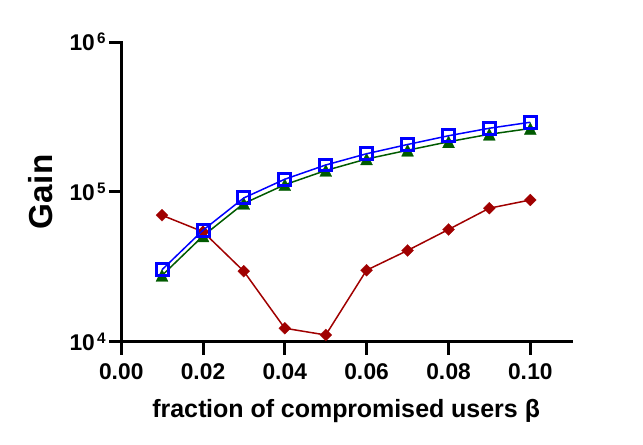}
& \includegraphics[width=0.2\textwidth]{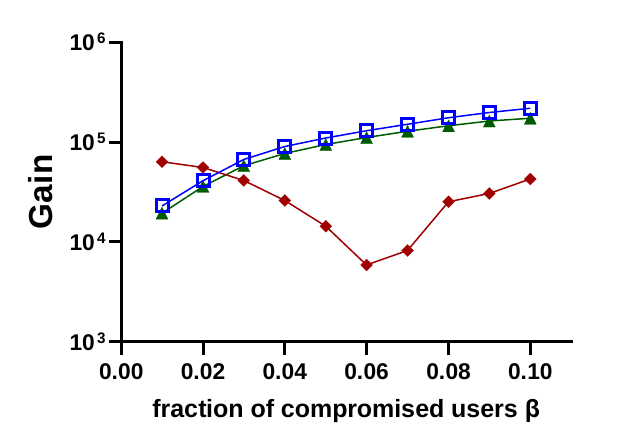}
& \includegraphics[width=0.2\textwidth]{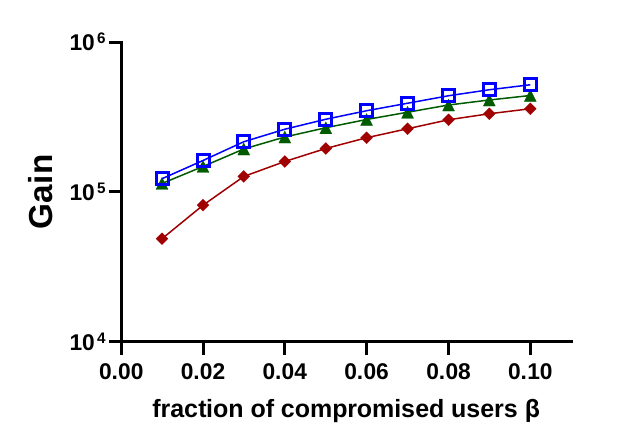}
\\
\graphSize{(a) LocalRR$_\triangle$, $\beta$}
& \graphSize{(b) TCountRR, $\beta$}
& \graphSize{(c) ARROneNS$_\triangle$, $\beta$}
& \graphSize{(d) TDP-VC, $\beta$}
& \graphSize{(e) EdgeOrient$_\triangle$, $\beta$}
\end{tabular}
\caption{Defenses against attacks on LDP triangle protocols  on ca-AstroPh with changing $\beta$ }
\label{fig:defense_triangle}
\end{figure*}

\section{Natural Defenses}
\label{sec:defense}

Existing robustness mechanisms for LDP graph data often rely on structural redundancy~\cite{roy2025robust,wang2024ldpdetect}. A common assumption is that an edge is reported by both incident users, so inconsistent reports can be used to identify suspicious participants. However, the LDP triangle counting protocols studied in this paper avoid such redundancy by collecting only upper-triangular entries, so each edge is reported once by a designated user. As a result, redundancy-based defenses cannot be directly applied. To assess whether simple aggregator-side mitigation is sufficient, we evaluate two natural post-processing defenses that attempt to detect suspicious users based on the perturbed reports received.

\subsection{Defense Strategies}

\myparatight{Prune}
This strategy assumes that compromised users tend to introduce many poisoned edges and therefore may become high-degree nodes in the perturbed graph. The aggregator marks the top $5\%$ nodes with the largest degrees in the perturbed graph as suspicious and removes their submitted reports from the estimation pipeline.

\myparatight{Frequent-pattern detection}
Inspired by frequent-pattern based detection~\cite{he2025data}, this strategy assumes that effective poisoning may create repeated structures across compromised reports. We treat each user's perturbed neighbor list as a transaction and each reported neighbor as an item, and then apply Apriori~\cite{srikant1996fast} to identify unusually frequent itemsets. A user is marked as suspicious if its report contains frequent patterns whose support exceeds a predefined threshold.

\myparatight{Mitigation rule}
After suspicious users are detected, their reports are excluded from graph reconstruction and triangle estimation. Since the considered protocols collect each upper-triangular edge only once, there is no redundant report from the other endpoint to recover the removed entry. Therefore, entries owned by marked users are treated as unavailable and are set to non-edges during reconstruction. This rule is intentionally simple and represents a natural aggregator-side post-processing defense, rather than a protocol-specific robust estimator.

\subsection{Performance of Detection Methods}

We evaluate the two defenses on ca-AstroPh under different fractions of compromised users. As shown in Fig.~\ref{fig:defense_triangle}, neither defense reliably mitigates DPA$_\triangle$. Prune is unstable because high-degree honest users can be mistakenly removed. Since such users often contribute substantially to the true triangle structure, excluding their reports may distort the clean LDP estimate and can even increase the absolute deviation from the undefended estimate. Frequent-pattern detection provides limited mitigation in some cases, but its effect remains weak when poisoned edges are distributed across different local structures rather than concentrated in repeated patterns.

These results highlight a key limitation of simple aggregator-side defenses. In LDP triangle counting, poisoned reports are protocol-compliant and statistically mixed with privacy noise, while the upper-triangular collection rule removes the redundancy needed for cross-user consistency checks. Therefore, detecting compromised users from perturbed reports alone is intrinsically difficult. More effective defenses likely need to be designed jointly with the reporting mechanism and the calibrated estimator, rather than added only as post-processing.

\section{Related Works}
\subsection{LDP Triangle Counting Protocols}
Differential privacy~\cite{dwork2006differential} has been widely studied for protecting sensitive graph-structured data.
Local differential privacy~\cite{wang2017locally}, which removes the need for a trusted curator, has been increasingly adopted in graph analytics, including synthetic graph generation~\cite{qin2017generating,wei2020asgldp}, graph structural statistic/metric estimation~\cite{ye2020lf}, graph data publication~\cite{xu2025continuous}, and subgraph counting~\cite{imola2021locally,imola2022communication,he2025robust,mundra2025practical}.

Our work focuses on LDP protocols for triangle counting. Imola et al.~\cite{imola2021locally} initiated a systematic study of subgraph counting under LDP and proposed representative non-interactive protocols such as LocalLap$_{k\star}$ and LocalRR$_\triangle$, with theoretical analysis for $k$-star and triangle counting.
They also introduced Local2Rounds$_\triangle$, an interactive two-round LDP protocol for triangle counting that leverages each user’s local view of the graph structure to improve estimation accuracy while reducing computational overhead. 
Building on this line, Imola et al.~\cite{imola2022communication} proposed communication-efficient interactive variants (e.g., ARRFull$_\triangle$, ARROneNS$_\triangle$, and ARRTwoNS$_\triangle$) by carefully designing the perturbed information broadcast by the aggregator. 
More recently, He et al.~\cite{he2025robust} proposed TDP-PC and TDP-VC, which incorporate sensitivity control and refined estimators to further improve accuracy.
Mundra et al. \cite{mundra2025practical} bound the sensitivity of triangle counting by leveraging a low-out-degree ordering computed via an LDP $k$-core protocol, thereby improving the accuracy of LDP triangle counting.

\subsection{Data Poisoning Attacks and Defenses on LDP Protocols}
Because LDP perturbation is performed locally and the aggregator never observes raw data, malicious reports are inherently difficult to distinguish from honest ones, which has motivated a growing body of work on data poisoning under LDP~\cite{cheu2021manipulation,cao2021data,wu2022poisoning,li2023fine,tong2024data,zhan2025poisoning,he2025data}.
Cao et al.~\cite{cao2021data} were among the first to systematically investigate data poisoning attacks against LDP protocols for frequency estimation and heavy hitter identification, and proposed maximal gain attacks that demonstrate the vulnerability of these two data analysis tasks under LDP. 
% Wu et al.~\cite{wu2022poisoning} formulated the attack objective as a double-objective optimization problem and generated poisoned data using approximate solutions, demonstrating the effectiveness of data poisoning attacks on key–value data. 
Tong et al.~\cite{tong2024data} introduced a unified attack framework with composable operators for poisoning LDP frequent itemset mining and demonstrated effectiveness under partial-knowledge settings. He et al.~\cite{he2025data} extended poisoning to graph analytics by proposing maximal gain attacks against LDP protocols for estimating graph metrics(e.g., degree centrality and clustering coefficient). 
% However, subgraph counting differs qualitatively from these settings: the attacker targets a \emph{global} combinatorial quantity, and the final output is often produced via calibration procedures tightly coupled with higher-order patterns. Our work addresses this gap and demonstrates the effectiveness of data poisoning attacks on LDP triangle counting tasks.

As poisoning threats to LDP protocols have expanded, a range of defenses has also been proposed~\cite{song2023efficient,sun2024ldprecover,huang2024ldpguard,wang2024ldp,horigome2024robust,li2025mitigating,zhao2025attack,roy2025robust,wang2024ldpdetect}.
These defenses include proof-based approaches (e.g., incorporating zero-knowledge proofs into interactive LDP protocols)~\cite{song2023efficient}, detection-and-mitigation frameworks such as LDPGuard~\cite{huang2024ldpguard}, and recovery-based post-processing methods~\cite{li2025mitigating}.
% Nevertheless, existing defenses are largely developed for frequency estimation and related settings, and they do not readily extend to LDP graph analytics where reports encode structured relational information.
Wang\cite{wang2024ldpdetect} and Roy\cite{roy2025robust} each proposed a defense strategy to mitigate data poisoning attacks in LDP graph data analysis. However, both approaches rely on edge redundancy. In contrast, the LDP triangle counting protocols considered in this paper avoid redundant reporting by collecting only the upper-triangular edges, which makes both defense strategies inapplicable.
This motivates the need for robustness mechanisms tailored to poisoning attacks in LDP-based subgraph counting.

\section{Conclusion}
To the best of our knowledge, this work presents the first systematic study of data poisoning attacks against LDP triangle counting. We develop a unified attack framework applicable to both non-interactive and interactive settings and instantiate it for representative LDP triangle counting protocols. We demonstrate the effectiveness of our attacks on four real-world datasets and validate them under multiple scenarios. In particular, our results show that the proposed attacks remain effective even under partial-knowledge settings. Our evaluation of natural defense mechanisms shows that existing LDP triangle counting protocols remain insufficiently robust, highlighting the need for more reliable designs that jointly consider privacy protection and adversarial robustness.

\section*{Acknowledgment}
This paper was edited for grammar using ChatGPT. 
\bibliographystyle{IEEEtran}
\bibliography{ref}

@inproceedings{cheu2021manipulation,
  title={Manipulation attacks in local differential privacy},
  author={Cheu, Albert and Smith, Adam and Ullman, Jonathan},
  booktitle={2021 IEEE Symposium on Security and Privacy (SP)},
  pages={883--900},
  year={2021},
  organization={IEEE}
}

@inproceedings{cao2021data,
  title={Data poisoning attacks to local differential privacy protocols},
  author={Cao, Xiaoyu and Jia, Jinyuan and Gong, Neil Zhenqiang},
  booktitle={30th USENIX Security Symposium (USENIX Security 21)},
  pages={947--964},
  year={2021}
}

@inproceedings{he2025data,
  title={Data Poisoning Attacks to Local Differential Privacy Protocols for Graphs},
  author={He, Xi and Huang, Kai and Ye, Qingqing and Hu, Haibo},
  booktitle={2025 IEEE 41st International Conference on Data Engineering (ICDE)},
  pages={987--1000},
  year={2025},
  organization={IEEE Computer Society}
}

@inproceedings{imola2021locally,
  title={Locally differentially private analysis of graph statistics},
  author={Imola, Jacob and Murakami, Takao and Chaudhuri, Kamalika},
  booktitle={30th USENIX security symposium (USENIX Security 21)},
  pages={983--1000},
  year={2021}
}

@inproceedings{zhao2025attack,
  title={An Attack-Agnostic Defense Framework Against Manipulation Attacks Under Local Differential Privacy},
  author={Zhao, Puning and Zhang, Zhikun and Dong, Jiawei and Wu, Jiafei and Liu, Zhe and Wang, Shaowei and Gao, Yunjun},
  booktitle={2025 IEEE Symposium on Security and Privacy (SP)},
  pages={3858--3876},
  year={2025},
  organization={IEEE}
}

@article{ye2020lf,
  title={LF-GDPR: A framework for estimating graph metrics with local differential privacy},
  author={Ye, Qingqing and Hu, Haibo and Au, Man Ho and Meng, Xiaofeng and Xiao, Xiaokui},
  journal={IEEE Transactions on Knowledge and Data Engineering},
  volume={34},
  number={10},
  pages={4905--4920},
  year={2020},
  publisher={IEEE}
}

@article{liu2024edge,
  title={Edge-protected triangle count estimation under relationship local differential privacy},
  author={Liu, Yuhan and Wang, Tianhao and Liu, Yixuan and Chen, Hong and Li, Cuiping},
  journal={IEEE Transactions on Knowledge and Data Engineering},
  volume={36},
  number={10},
  pages={5138--5152},
  year={2024},
  publisher={IEEE}
}

@inproceedings{imola2022communication,
  title={$\{$Communication-Efficient$\}$ triangle counting under local differential privacy},
  author={Imola, Jacob and Murakami, Takao and Chaudhuri, Kamalika},
  booktitle={31st USENIX security symposium (USENIX Security 22)},
  pages={537--554},
  year={2022}
}

@article{dehghani2015using,
  title={Using shortcut edges to maximize the number of triangles in graphs},
  author={Dehghani, Sina and Fazli, Mohammad Amin and Habibi, Jafar and Yazdanbod, Sadra},
  journal={Operations Research Letters},
  volume={43},
  number={6},
  pages={586--591},
  year={2015},
  publisher={Elsevier}
}

@inproceedings{dwork2006differential,
  title={Differential privacy},
  author={Dwork, Cynthia},
  booktitle={International colloquium on automata, languages, and programming},
  pages={1--12},
  year={2006},
  organization={Springer}
}

@inproceedings{wang2017locally,
  title={Locally differentially private protocols for frequency estimation},
  author={Wang, Tianhao and Blocki, Jeremiah and Li, Ninghui and Jha, Somesh},
  booktitle={26th USENIX security symposium (USENIX Security 17)},
  pages={729--745},
  year={2017}
}

@inproceedings{ye2019privkv,
  title={PrivKV: Key-value data collection with local differential privacy},
  author={Ye, Qingqing and Hu, Haibo and Meng, Xiaofeng and Zheng, Huadi},
  booktitle={2019 IEEE Symposium on Security and Privacy (SP)},
  pages={317--331},
  year={2019},
  organization={IEEE}
}

@article{errounda2021collective,
  title={Collective location statistics release with local differential privacy},
  author={Errounda, Fatima Zahra and Liu, Yan},
  journal={Future Generation Computer Systems},
  volume={124},
  pages={174--186},
  year={2021},
  publisher={Elsevier}
}

@inproceedings{erlingsson2014rappor,
  title={Rappor: Randomized aggregatable privacy-preserving ordinal response},
  author={Erlingsson, {\'U}lfar and Pihur, Vasyl and Korolova, Aleksandra},
  booktitle={Proceedings of the 2014 ACM SIGSAC conference on computer and communications security},
  pages={1054--1067},
  year={2014}
}

@inproceedings{kasiviswanathan2013analyzing,
  title={Analyzing graphs with node differential privacy},
  author={Kasiviswanathan, Shiva Prasad and Nissim, Kobbi and Raskhodnikova, Sofya and Smith, Adam},
  booktitle={Theory of Cryptography Conference},
  pages={457--476},
  year={2013},
  organization={Springer}
}

@inproceedings{qin2017generating,
  title={Generating synthetic decentralized social graphs with local differential privacy},
  author={Qin, Zhan and Yu, Ting and Yang, Yin and Khalil, Issa and Xiao, Xiaokui and Ren, Kui},
  booktitle={Proceedings of the 2017 ACM SIGSAC conference on computer and communications security},
  pages={425--438},
  year={2017}
}

@article{he2025robust,
  title={Robust Privacy-Preserving Triangle Counting under Edge Local Differential Privacy},
  author={He, Yizhang and Wang, Kai and Zhang, Wenjie and Lin, Xuemin and Zhang, Ying and Ni, Wei},
  journal={Proceedings of the ACM on Management of Data},
  volume={3},
  number={3},
  pages={1--26},
  year={2025},
  publisher={ACM New York, NY, USA}
}

@misc{snapnets,
    author= {Jure Leskovec and Andrej Krevl},
    title= {{SNAP Datasets}: {Stanford} Large Network Dataset Collection},
    howpublished = {\url{http://snap.stanford.edu/data}},
    mon= jun,
    year= 2014
}

@article{horigome2024robust,
  title={Robust Estimation Method against Poisoning Attacks for Key-Value Data with Local Differential Privacy},
  author={Horigome, Hikaru and Kikuchi, Hiroaki and Fujita, Masahiro and Yu, Chia-Mu},
  journal={Applied Sciences},
  volume={14},
  number={14},
  pages={6368},
  year={2024},
  publisher={MDPI}
}

@book{srikant1996fast,
  title={Fast algorithms for mining association rules and sequential patterns},
  author={Srikant, Ramakrishnan},
  year={1996},
  publisher={The University of Wisconsin-Madison}
}

@inproceedings{tong2024data,
  title={Data poisoning attacks to locally differentially private frequent itemset mining protocols},
  author={Tong, Wei and Chen, Haoyu and Niu, Jiacheng and Zhong, Sheng},
  booktitle={Proceedings of the 2024 on ACM SIGSAC Conference on Computer and Communications Security},
  pages={3555--3569},
  year={2024}
}

@article{wei2020asgldp,
  title={AsgLDP: Collecting and generating decentralized attributed graphs with local differential privacy},
  author={Wei, Chengkun and Ji, Shouling and Liu, Changchang and Chen, Wenzhi and Wang, Ting},
  journal={IEEE Transactions on Information Forensics and Security},
  volume={15},
  pages={3239--3254},
  year={2020},
  publisher={IEEE}
}

@article{xu2025continuous,
  title={Continuous Publication of Weighted Graphs with Local Differential Privacy},
  author={Xu, Wen and Qiao, Pengpeng and Liu, Shang and Zheng, Zhirun and Cao, Yang and Li, Zhetao},
  journal={Proceedings of the VLDB Endowment},
  volume={18},
  number={11},
  pages={4214--4226},
  year={2025},
  publisher={VLDB Endowment}
}

@inproceedings{zhan2025poisoning,
  title={Poisoning Attacks to Local Differential Privacy for Ranking Estimation},
  author={Zhan, Pei and Tang, Peng and Li, Yangzhuo and Wei, Puwen and Guo, Shanqing},
  booktitle={Proceedings of the 2025 ACM SIGSAC Conference on Computer and Communications Security},
  pages={1215--1229},
  year={2025}
}

@inproceedings{wu2022poisoning,
  title={Poisoning attacks to local differential privacy protocols for $\{$Key-Value$\}$ data},
  author={Wu, Yongji and Cao, Xiaoyu and Jia, Jinyuan and Gong, Neil Zhenqiang},
  booktitle={31st USENIX Security Symposium (USENIX Security 22)},
  pages={519--536},
  year={2022}
}

@inproceedings{li2023fine,
  title={Fine-grained poisoning attack to local differential privacy protocols for mean and variance estimation},
  author={Li, Xiaoguang and Li, Ninghui and Sun, Wenhai and Gong, Neil Zhenqiang and Li, Hui},
  booktitle={32nd USENIX Security Symposium (USENIX Security 23)},
  pages={1739--1756},
  year={2023}
}

@article{huang2024ldpguard,
  title={LDPGuard: Defenses against data poisoning attacks to local differential privacy protocols},
  author={Huang, Kai and Ouyang, Gaoya and Ye, Qingqing and Hu, Haibo and Zheng, Bolong and Zhao, Xi and Zhang, Ruiyuan and Zhou, Xiaofang},
  journal={IEEE Transactions on Knowledge and Data Engineering},
  volume={36},
  number={7},
  pages={3195--3209},
  year={2024},
  publisher={IEEE}
}

@inproceedings{li2025mitigating,
  title={Mitigating Data Poisoning Attacks to Local Differential Privacy},
  author={Li, Xiaolin and Li, Ninghui and Wang, Boyang and Sun, Wenhai},
  booktitle={Proceedings of the 2025 ACM SIGSAC Conference on Computer and Communications Security},
  pages={1230--1244},
  year={2025}
}

@inproceedings{sun2024ldprecover,
  title={Ldprecover: Recovering frequencies from poisoning attacks against local differential privacy},
  author={Sun, Xinyue and Ye, Qingqing and Hu, Haibo and Duan, Jiawei and Wo, Tianyu and Xu, Jie and Yang, Renyu},
  booktitle={2024 IEEE 40th International Conference on Data Engineering (ICDE)},
  pages={1619--1631},
  year={2024},
  organization={IEEE}
}

@article{song2023efficient,
  title={Efficient defenses against output poisoning attacks on local differential privacy},
  author={Song, Shaorui and Xu, Lei and Zhu, Liehuang},
  journal={IEEE Transactions on Information Forensics and Security},
  volume={18},
  pages={5506--5521},
  year={2023},
  publisher={IEEE}
}

@inproceedings{wang2024ldp,
  title={LDP-Purifier: Defending against poisoning attacks in local differential privacy},
  author={Wang, Leixia and Ye, Qingqing and Hu, Haibo and Meng, Xiaofeng and Huang, Kai},
  booktitle={International Conference on Database Systems for Advanced Applications},
  pages={221--231},
  year={2024},
  organization={Springer}
}

@inproceedings{horigome2023local,
  title={Local differential privacy protocol for making key--value data robust against poisoning attacks},
  author={Horigome, Hikaru and Kikuchi, Hiroaki and Yu, Chia-Mu},
  booktitle={International Conference on Modeling Decisions for Artificial Intelligence},
  pages={241--252},
  year={2023},
  organization={Springer}
}

@article{li2024robustness,
  title={On the robustness of ldp protocols for numerical attributes under data poisoning attacks},
  author={Li, Xiaoguang and Li, Zitao and Li, Ninghui and Sun, Wenhai},
  journal={arXiv preprint arXiv:2403.19510},
  year={2024}
}

@inproceedings{wang2019attacking,
  title={Attacking graph-based classification via manipulating the graph structure},
  author={Wang, Binghui and Gong, Neil Zhenqiang},
  booktitle={Proceedings of the 2019 ACM SIGSAC conference on computer and communications security},
  pages={2023--2040},
  year={2019}
}

@inproceedings{cheng2025atom,
  title={Atom: A framework of detecting query-based model extraction attacks for graph neural networks},
  author={Cheng, Zhan and Shen, Bolin and Sha, Tianming and Gao, Yuan and Li, Shibo and Dong, Yushun},
  booktitle={Proceedings of the 31st ACM SIGKDD Conference on Knowledge Discovery and Data Mining V. 2},
  pages={322--333},
  year={2025}
}

@inproceedings{wu2022model,
  title={Model extraction attacks on graph neural networks: Taxonomy and realisation},
  author={Wu, Bang and Yang, Xiangwen and Pan, Shirui and Yuan, Xingliang},
  booktitle={Proceedings of the 2022 ACM on Asia conference on computer and communications security},
  pages={337--350},
  year={2022}
}

@article{mundra2025practical,
  title={Practical and Accurate Local Edge Differentially Private Graph Algorithms},
  author={Mundra, Pranay and Papamanthou, Charalampos and Shun, Julian and Liu, Quanquan C},
  journal={Proceedings of the VLDB Endowment},
  volume={18},
  number={11},
  pages={4199--4213},
  year={2025},
  publisher={VLDB Endowment}
}

@article{eden2025triangle,
  title={Triangle counting with local edge differential privacy},
  author={Eden, Talya and Liu, Quanquan C and Raskhodnikova, Sofya and Smith, Adam},
  journal={Random Structures \& Algorithms},
  volume={66},
  number={4},
  pages={e70002},
  year={2025},
  publisher={Wiley Online Library}
}

@inproceedings{roy2025robust,
  title={Robust locally differentially private graph analysis},
  author={Roy Chowdhury, Amrita and Imola, Jacob and Chaudhuri, Kamalika},
  booktitle={Proceedings of the 20th ACM Asia Conference on Computer and Communications Security},
  pages={635--650},
  year={2025}
}

@inproceedings{wang2024ldpdetect,
  title={LDPDetect: Efficient Detection and Defense Against Graph Poisoning Attacks Under Local Differential Privacy},
  author={Wang, Teng and Liang, Jiangguo and Zhao, Ling and Hu, Zhi},
  booktitle={International Conference on Frontiers in Cyber Security},
  pages={53--69},
  year={2024},
  organization={Springer}
}

@article{warner1965randomized,
  title={Randomized response: A survey technique for eliminating evasive answer bias},
  author={Warner, Stanley L},
  journal={Journal of the American statistical association},
  volume={60},
  number={309},
  pages={63--69},
  year={1965},
  publisher={Taylor \& Francis}
}

@article{kairouz2014extremal,
  title={Extremal mechanisms for local differential privacy},
  author={Kairouz, Peter and Oh, Sewoong and Viswanath, Pramod},
  journal={Advances in neural information processing systems},
  volume={27},
  year={2014}
}

\end{document}